\documentclass[sigconf,edbt]{acmart-edbt2021}

\def\BibTeX{{\rm B\kern-.05em{\sc i\kern-.025em b}\kern-.08em
    T\kern-.1667em\lower.7ex\hbox{E}\kern-.125emX}}

\usepackage{booktabs} 

\setcopyright{none}
\renewcommand\footnotetextcopyrightpermission[1]{}
\usepackage{graphicx}
\usepackage{pbox}
\usepackage{multirow}
\usepackage{color}
\usepackage{balance}  
\usepackage{algorithm, algorithmicx}
\usepackage{algpseudocode}

\usepackage{enumitem}
\usepackage{subcaption}

\newcommand{\plam}{p_{\lambda}}
\newcommand{\clam}{C_{\lambda}}
\newcommand{\hlam}{HAC($\lambda$)}
\newcommand{\alphalam}{\alpha_{\lambda}}

\newcommand{\mathv}{\mathcal{V}}

\newcommand{\rof}{\rho_f}
\newcommand{\ros}{\rho_s}
\newcommand{\rop}{\rho_p}
\newcommand{\rod}{\rho_d}
\newcommand{\rom}{\rho_m}
\newcommand{\roz}{\rho_z}
\newcommand{\ror}{\rho_r}
\newcommand{\nlam}{n_{\lambda}}
\newcommand{\rlam}{r_{\lambda}}

\newcommand{\manual}{{\sf Manual}}
\newcommand{\quack}{{\sf Quack}}

\newcommand{\cc}{{\sf Winston}}
\newcommand{\clang}{{\sf Winston}}
\newcommand{\winston}{{\sf Winston}}
\newcommand{\orefine}{{\sf OpenRefine}}
\newcommand{\transer}{{\sf TransER}}
\newcommand{\falcon}{{\sf Falcon}}
\newcommand{\magellan}{{\sf Magellan}}
\newcommand{\waldo}{{\sf Waldo}}

\renewcommand{\paragraph}[1]{\vspace{0.5mm}\noindent{\bf #1}\ \ }

\newcommand{\note}[1]{{\color{red} #1}}
\newcommand{\blue}[1]{{\color{blue} #1}}

\newcommand{\bfwinston}{{\sf \textbf{Winston}}}
\newcommand{\cwinston}{{\sf cWinston}}
\newcommand{\bcwinston}{{\sf \textbf{cWinston}}}

\newcommand{\mysplit}{{\sf Split}}
\newcommand{\splitcluster}{{\sf SplitCluster}}
\newcommand{\lmerge}{{\sf LocalMerge}}
\newcommand{\gmerge}{{\sf GlobalMerge}}
\newcommand{\merge}{{\sf Merge}}

\usepackage{etoolbox}
\newtoggle{adel}

\newlength{\algindent}
\newcommand{\Indstate}[1][1]{\State\hspace{#1\algindent}}

\newcommand{\uafocus}{\underline{\texttt{focus}}}

\newcommand{\uamatch}{\underline{\texttt{match}}}
\newcommand{\uaispure}{\underline{\texttt{is}}\-\underline{\texttt{Pure}}}
\newcommand{\uaselect}{\underline{\texttt{select}}}
\newcommand{\uafinddom}{\underline{\texttt{find}}\-\underline{\texttt{Dom}}\-\underline{\texttt{\smash{Entity}}}\-\underline{\texttt{Value}}}

\newcommand{\uamemorize}{\underline{\texttt{memorize}}}
\newcommand{\uarecall}{\underline{\texttt{recall}}}

\newcommand{\albr}{\allowbreak}

\toggletrue{adel}

\begin{document}
	
\title{Toward Data Cleaning with a Target Accuracy:\\
A Case Study for Value Normalization}

\author{Adel Ardalan}
\affiliation{%
	\institution{University of Wisconsin-Madison}
}
\email{adel@cs.wisc.edu}

\author{Derek Paulsen}
\affiliation{%
	\institution{University of Wisconsin-Madison}
}
\email{dpaulsen2@wisc.edu}

\author{Amanpreet Singh Saini}
\affiliation{%
	\institution{University of Wisconsin-Madison}
}
\email{saini5@wisc.edu}

\author{Walter Cai}
\affiliation{%
	\institution{University of Washington}
}
\email{walter@cs.washington.edu}

\author{AnHai Doan} 
\affiliation{%
	\institution{University of Wisconsin-Madison}
}
\email{anhai@cs.wisc.edu}

\renewcommand{\shortauthors}{}

\begin{abstract}
Many applications need to clean data {\em with a target accuracy}.
As far as we know, this problem has not been studied in depth. In
this paper we take the first step toward solving it. We focus on
{\em value normalization (VN)}, the problem of replacing all string
that refer to the same entity with a unique string.  VN is
ubiquitous, and we often want to do VN with 100\% accuracy. This is
typically done today in industry by automatically clustering the
strings then asking a user to verify and clean the clusters, until
reaching 100\% accuracy. This solution has significant
limitations. It does not tell the users how to verify and clean the
clusters. This part also often takes a lot of time, e.g.,
days. Further, there is no effective way for multiple users to
collaboratively verify and clean.  In this paper we address these
challenges. Overall, our work advances the state of the art in data
cleaning by introducing a novel cleaning problem and describing a
promising solution template.
\end{abstract}

\maketitle

\section{Introduction}\label{intro}

Data cleaning (DC) has been a long-standing challenge in the database
community. Many DC problems have been studied, such as cleaning with a
budget, cleaning to satisfy constraints but minimize changes to the data,
etc. Recently, however, we have seen another novel DC
problem in industry: {\em cleaning with a target accuracy}, e.g., with at
least 95\% precision and 90\% recall.  While pervasive, this problem
appears to have received little attention in the research community.

\begin{figure}
	\centering\includegraphics[height=1.1in]{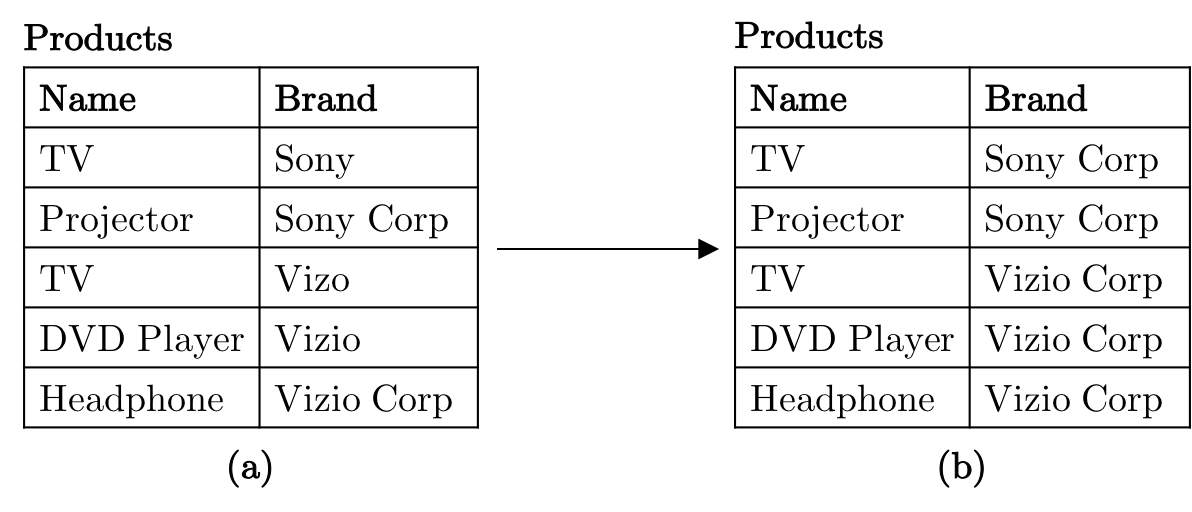} 
	\vspace{-4mm}\caption{An example of normalizing product brands}\label{ex1}
	\vspace{-1mm}
\end{figure}

In this work, we take the first step toward solving this problem.  We
focus on {\em value normalization (VN)}, the problem of replacing all
strings (in a given set) that refer to the same real-world entity with a
unique string. VN is ubiquitous, and industrial users often want to do
VN with 100\% accuracy.
\begin{example}{\em 
		To enable product browsing by brand on walmart.com, the business
		group at WalmartLabs asks the IT group to normalize the brands,
		e.g., converting
		those in Figure \ref{ex1}.a into those in Figure \ref{ex1}.b. If some
		brands are not normalized correctly, then customers may not find those products,
		resulting in revenue losses. So the business group
		asks the IT group to ensure that the brands are normalized with 100\% accuracy.}
\end{example}
Many enterprise customers of Informatica
(which sells data integration software) also face
this problem, in building business glossaries, master
data management, and knowledge graph construction. In general, {\em if even a small amount
	of inaccuracy in VN can cause significant problems for the target application,
	then the business group will typically ask the IT group to help perform VN
	with 100\% accuracy.}

\begin{figure}[b]
	\centering
	\hspace*{-4.25mm}
	\includegraphics[width=3.5in]{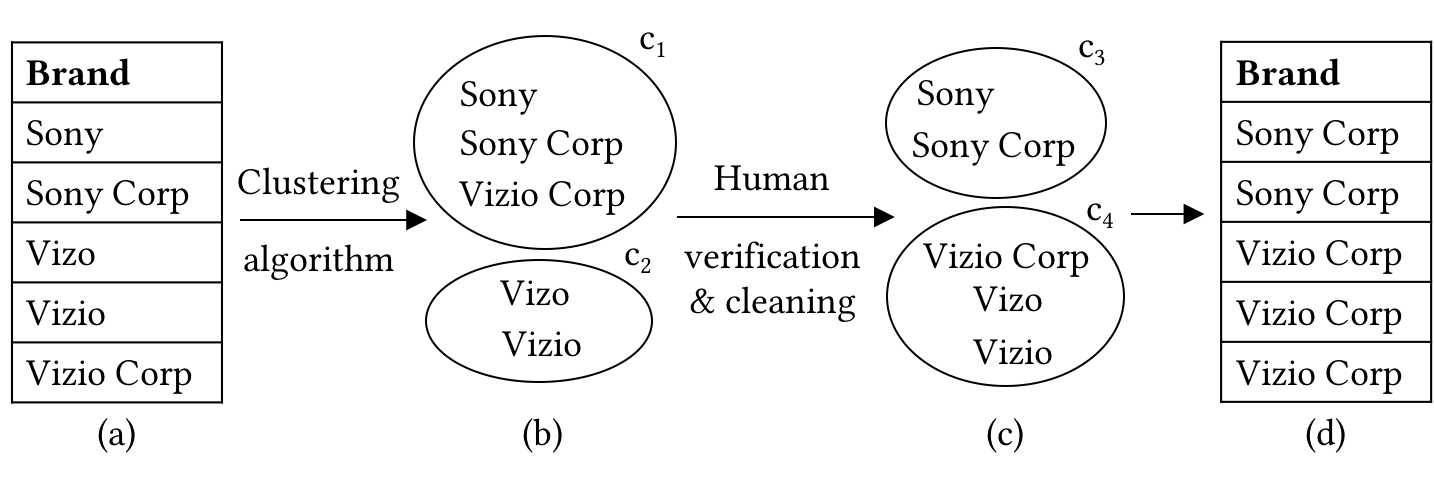}
	\vspace{-7mm}
	\caption{A popular solution in industry to perform VN with 100\% accuracy.}\label{ex2}
\end{figure}

In response, the IT group typically employs an algorithm to cluster
the strings, then asks a user to verify and clean the clusters. 
Consider the five brands in Figure \ref{ex2}.a. The IT group applies
an algorithm to produce two clusters $c_1$ and $c_2$ (Figure
\ref{ex2}.b). A user $U$ manually verifies and cleans the
clusters, by moving ``Vizio Corp'' from cluster $c_1$ to $c_2$, producing the
two clusters $c_3$ and $c_4$ in Figure \ref{ex2}.c. Finally, $U$ replaces
each string in a cluster with a canonical string, producing the VN
result in Figure \ref{ex2}.d.

Typically, a data scientist performs {\em the ``machine'' part\/} that
clusters the strings, then a data analyst performs {\em the
	``human'' part\/} that verifies and cleans the clusters. The IT group
assures the business group that the resulting output is 100\% accurate
because a data analyst has examined it (assuming that he/she does
not make mistakes).

While popular, the above solution has significant limitations. First,
there is no precise procedure that a user can follow to execute the
``human'' part. So users often verify and clean in an 
ad-hoc, suboptimal, and often incorrect fashion.  This also
makes it impossible to understand the assumptions under which the
solution reaches 100\% accuracy and to formally prove it. 

Second, the ``human'' part often incurs a huge amount of time, e.g.,
days. In contrast, the ``machine'' part often takes mere minutes. (In
most cases that we have seen, users verified and cleaned using Excel,
in a slow and tedious process.) So it is critical to develop a better
solution and GUI tool to minimize the time of the ``human''
part.

Finally, it is difficult for multiple users to collaboratively verify
and clean, even though this setting commonly occurs in
practice.

In this paper we develop \winston, a solution for the above challenges.
 (In the movie ``Pulp Fiction'', Winston Wolfe is the fixer who cleans up messes made by other gangsters.) 
We first define a set
of basic operations on a GUI for users, e.g., selecting a value,
verifying if a cluster is clean, merging two clusters, etc. Then we
provide {\em precise procedures\/} involving these actions that users can
execute to verify/clean clusters. We prove that if users execute these
actions correctly, then the output has 100\% accuracy.

To minimize the time of the ``human'' part, we adopt an RDBMS-style
solution. Specifically, we
compose the GUI operations with clustering algorithms to form multiple
``machine-human'' plans, each executes the VN pipeline
end-to-end. Next, we estimate the total time a user must spend per
plan, select the plan with the least estimated time, execute its
machine part, then show the output of that part to the user so that
he/she can verify and clean it using a GUI (following the sequence of
user operations that this plan specifies).

Finally, we show how to extend our solution to effectively divide the
verification and cleaning work among multiple users.

Our solution appears highly effective. Section \ref{eval} shows that
using the existing solution, a single user
needs 29 days, 4.4 years, and 11.5 years to verify/clean 100K, 500K, and 1M
strings, respectively. \winston\ drastically reduces these times
to just 13 days, 9.6 months, and 1.3 years, using 1 user, and to 
4.25 days, 2.2 months, and 3.5 months, using 3 users.

In summary, we make the following contributions:
\begin{itemize}
	\item We formally define the novel data cleaning problem
	of VN with 100\% accuracy. As far as we know, this paper is the
	first to study this problem in depth.
	
	\item We propose \cc, a novel RDBMS-style solution. \cc\ defines 
	complex human operations and optimizes the human time of a
	plan. This is in contrast to traditional RDBMSs which define machine
	operations and optimize machine time.
	
	\item We describe extensive experiments (comparing \cc\ to a
	tool in a company, to the popular open-source tool
	\orefine, and to state-of-the-art string matching and entity
	matching solutions) that show the promise of our approach.
\end{itemize}
Overall, our work {\em advances the state of the art in data cleaning\/} by
introducing a novel cleaning problem and describing a promising solution template.
It also {\em advances the state of the art in human-in-the-loop data analytics (HILDA)\/}
by showing that it is possible to develop an RDBMS-style solution to HILDA,
by defining complex human operations, combining them to form plans, and selecting
the plan with the lowest estimated human effort.

\vspace{0mm}\section{Problem Definition}

In this section we define the problem of VN with 100\% accuracy, and examine
when we can reach this accuracy under what conditions. We first define
\vspace*{-1mm}\begin{definition}[Value normalization]\label{def1}
	Let $V$ be a set of strings $\{v_1, \ldots, v_n\}$. Replace each
	$v\in V$ with a string $s(v)$ such that $s(v_i) = s(v_j)$ if and
	only if $v_i$ and $v_j$ refer to the same real-world entity, for all
	$v_i, v_j$ in $V$.
\end{definition}
This problem is often solved in two steps: (1) partition $V$
into a set of {\em disjoint\/} clusters $\mathv = \{V_1, \ldots,
V_m\}$, such that two strings refer to the same real-world entity if
and only if they belong to the same cluster; (2) replace all
strings in each cluster $V_i\in \mathv$ with a canonical string $s_i$.

In this paper we will consider only Step 1, which tries to find the
correct partitioning of $V$. Step 2 is typically application dependent
(e.g., a common method is to select the longest string in a cluster
$V_i$ to be its canonical string, because this string tends to be the
most informative one).

\paragraph{Gold Partition \& Accuracy of Partitions:} Let $U$ be
a user who will verify and clean the clusters. To do so, $U$ must be
capable of creating a ``gold'', i.e., correct, partition $\mathv^* =
\{V_1^*, \ldots, V_k^*\}$, such that two strings refer to the same
real-world entity if and only if they are in the same cluster. For
example, the two clusters $c_3$ and $c_4$ in Figure \ref{ex2}.c form
the gold partition for the set of strings in Figure \ref{ex2}.a.

Our goal is to find the gold partition $\mathv^*$. But the
partition that we find may not be as accurate. We now
describe how to compute the accuracy of any partition. First,
we define
\begin{definition}[Match and non-match]
	A match $v_i = v_j$ means ``$v_i$ and $v_j$ refer to the
	same real-world entity'', and is correct if this is indeed
	true. $v_i = v_j$ and $v_j = v_i$ are considered the same match. We
	define a non-match $v_i \ne v_j$ similarly.
\end{definition}
\begin{definition}[Set of matches specified by a partition]
	A cluster $V_i$ specifies the set of matches $M(V_i) = \{v_p = v_q| v_p\in V_i, v_q\in V_i, p\ne q\}$. Partition $\mathv = \{V_1,
	\ldots, V_m\}$ specifies the set of matches $M(\mathv) = \cup_{i=1}^m M(V_i)$.
\end{definition}
For example, cluster $c_1$ in Figure \ref{ex2}.b specifies three
matches: $M(c_1) = \{Sony = Sony\ Corp, Sony = Vizio\ Corp, Sony\ Corp
= Vizio\ Corp\}$. Cluster $c_2$ specifies one match: $M(c_2) = \{Vizio
= Vizio\ Inc\}$. These two clusters form a partition $\mathv_{12}$,
which specifies the set of matches $M(c_1) \cup M(c_2)$.  The accuracy
of a partition is then measured as follows:
\begin{definition}[Precision and recall of a partition]
	Let $\mathv^*$ be the gold partition of a set of strings $V$. The
	precision of a partition $\mathv$ is the
	fraction of matches in $M(\mathv)$ that are correct, i.e., appearing
	in $M(\mathv^*)$, and the recall of $\mathv$ is the fraction of
	matches in $M(\mathv^*)$ that appear in $M(\mathv)$.
\end{definition}
Given the gold partition $\mathv_{34} = \{c_3, c_4\}$ in Figure
\ref{ex2}.c, the precision of partition $\mathv_{12} = \{c_1, c_2\}$ in
Figure \ref{ex2}.b is 2/4 = 50\%, and the recall is 2/4 = 50\%.

\begin{figure}
	\centering
	\includegraphics[height=1.3in]{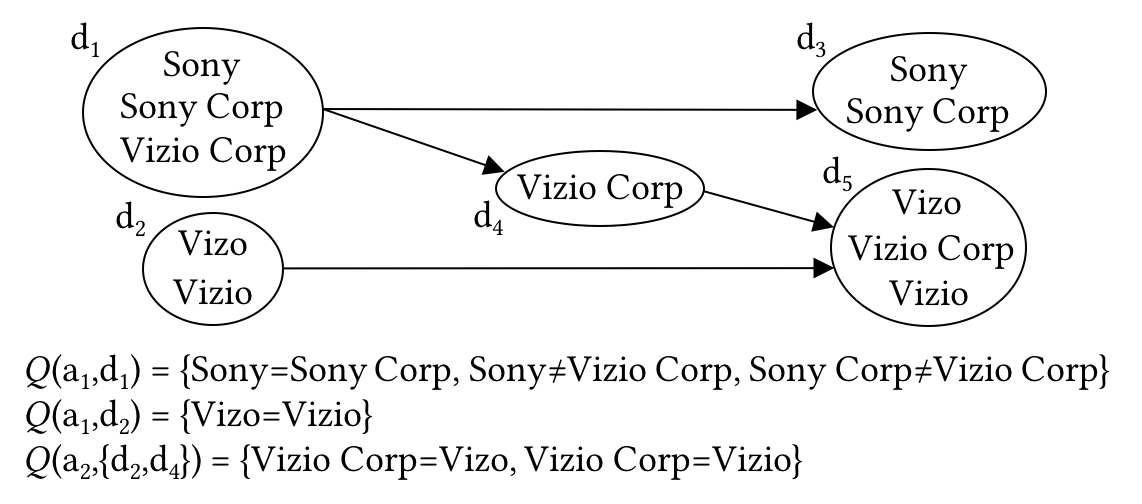} 
	\vspace{-3mm}\caption{Actions and their verification 
		sets}\label{ex3}
	\vspace{-2mm}
\end{figure}

\paragraph{Actions \& Their Verification Sets:} Henceforth,
we use ``action'' and ``operation'' interchangeably. When a user $U$ performs
an action, $U$ has implicitly verified a set of matches and non-matches,
called {\em a verification set}. Formally,
\begin{definition}[User action and verification set]
	We assume a GUI on which user $U$ can
	perform a set of actions $A = \{a_1, \ldots, a_r\}$. Each action
	$a_i$ inputs data $I_i$ and outputs data $O_i$, both of which
	involve sets of strings in $V$. After correctly executing an action $a_i$
	on input $I_i$, as a side effect, user $U$ has implicitly verified
	a set $Q(a_i, I_i)$ of matches and non-matches to be correct. We refer to
	$Q(a_i, I_i)$ as a verification set.
\end{definition}
To illustrate, suppose $U$ has employed an algorithm to produce
the partition $\{d_1, d_2\}$ in Figure \ref{ex3}. Next, $U$ uses a GUI
to verify and clean these clusters. Call a cluster ``pure'' if all
strings in it refer to the same real-world entity. Suppose the GUI
supports only two actions:  $a_1$ splits a cluster into pure
clusters, and $a_2$ merges two pure clusters into one.

Suppose $U$ starts by using $a_1$ to split cluster $d_1$
into two pure clusters $d_3 = \{Sony, Sony\ Corp\}$ and $d_4 =
\{Vizio\ Corp\}$ (see Figure \ref{ex3}). Cluster $d_1$ specifies three
matches: $Sony = Sony\ Corp$, $Sony = Vizio\ Corp$, and $Sony\ Corp =
Vizio\ Corp$. With the above splitting, intuitively, $U$ has
verified that the first match $Sony = Sony\ Corp$ is indeed correct,
but the remaining two matches are not. Thus, the resulting
verification set $Q(a_1, d_1)$ is the set of 1 match and 2
non-matches: $\{Sony = Sony\ Corp, Sony \ne Vizio\ Corp,$ $Sony\ Corp
\ne Vizio\ Corp\}$, as shown in Figure \ref{ex3}.

Next, $U$ uses $a_1$ to split cluster $d_2$. $U$ determines that $d_2$
is already pure, so no new clusters are created. Implicitly, $U$ has
verified that the sole match specified by $d_2$ is correct. So
$Q(a_1,d_2) = \{Vizo = Vizio\}$. Finally, $U$ uses action $a_2$ to
merge the two pure clusters $d_2$ and $d_4$ into cluster $d_5$ (see
Figure \ref{ex3}). Implicitly, $U$ has verified that the two matches
$\{Vizo = Vizio\ Corp, Vizio = Vizio\ Corp\}$ are correct. These form
the verification set $Q(a_2, \{d_2, d_4\})$.

In a similar fashion, we can define
the verification set for the sequence $a_1(d_1), a_1(d_2),
a_2(\{d_2,d_4\})$ in Figure \ref{ex3} to be $Q(a_1,d_1)\cup Q(a_1,d_2)\cup Q(a_2, \{d_2, d_4\})$.
Formally,
\vspace*{-1mm}\begin{definition}[Verification set of action sequence] If
	user $U$ has performed an action sequence $G = a_{i1}, \ldots,
	a_{il}$, then the verification set of $G$ is $Q(G) = \cup_{q=1}^l  Q(a_{iq})$.
\end{definition}

\paragraph{Match Transitivity:} Recall that we assume user $U$ can create a 
gold partition $\mathv^*$. This implies that transitivity holds for
matches, i.e., if $v_i = v_j$ and $v_j = v_h$ are correct, then $v_i =
v_h$ is also correct (because all three must be in the same gold
cluster). Similarly, if $v_i = v_j$ and $v_j \ne v_h$ are correct,
then $v_i \ne v_h$ is also correct.
\begin{definition}[Inferring matches]
	We say that match $v_i = v_j$ can be inferred from a verification set
	$Q(G)$ if and only if there exists a sequence of strings $v_{h1},
	\ldots, v_{hl}$ such that the matches $v_i = v_{h1}, v_{h1} = v_{h2},\ldots, v_{hl} = v_j$ are in $Q(G)$. We say that these matches form a
	transitivity path from $v_i$ to $v_j$. Similarly, we say
	that non-match $v_i \ne v_j$ can be inferred from $Q(G)$ if and
	only if there exists such a path, except that exactly one of the
	edges of the path is a non-match.
\end{definition}

\paragraph{VN with 100\% Accuracy:} Recall that the gold
partition $\mathv^*$ specifies a set of correct matches $M(\mathv^*)$.
We say that it also specifies a set of correct non-matches $N(\mathv^*)$,
which consists of all non-match $v_i \ne v_j$ such that $v_i = v_j$ is not
a match in $M(\mathv^*)$. We define
\begin{definition}[Gold sequence of actions]
	A sequence $G$ of actions of user $U$ is ``gold'' if and only if any
	match in $M(\mathv^*)$ or non-match in $N(\mathv^*)$ either already
	exists in the verification set $Q(G)$ or can be inferred from $Q(G)$.
\end{definition}
It is not difficult to prove that executing a gold action sequence $G$
will produce the gold partition $\mathv^*$. We now can define our
problem as follows:
\begin{definition}[VN with 100\% accuracy]
	Let $V$ be a set of strings. Let $(X,Y)$ be a pair of machine/human algorithms, such
	that the machine part $X$ can be executed on $V$ to produce a partition $\mathv$, then the human part $Y$ can be
	executed by a user $U$ on $\mathv$ to produce a new partition $\mathcal{V}+$. Find $X$
	and $Y$ such that (a) the action sequence executed by user $U$ in part $Y$ is a gold sequence, and (b)
	the total time spent by user $U$ is minimized. Return the resulting partition $\mathcal{V}+$. 
\end{definition}
Thus, we reach 100\% accuracy if the user executes a {\em gold\/}
sequence $G$ of actions. Then all correct matches and non-matches will
have already been in the verification set of $G$, or inferred from
this verification set via match transitivity.

\vspace{-1mm}\section{Defining the Human Part}\label{secplans}

As discussed, each VN plan $(X,Y)$ consists of a machine part $X$ and
a human part $Y$.  In part $X$ we apply an algorithm to the input
strings to obtain a set of clusters $\mathv$, then in part $Y$ we
employ a user $U$ to verify and clean $\mathv$. We now design part
$Y$; the next section designs part $X$.

The key challenge in designing the human part $Y$ is to ensure that it is
easy for users to understand and execute, minimizes their effort, and
is amenable to cost analysis. Toward these goals, we discuss the user
setting, describe a solution called {\em split and merge}, then define
a set of user operations that can be used to implement this solution.

We assume user $U$ will work with a graphical user interface (GUI),
using mouse and keyboard. $U$ has a short-term memory (or {\em STM\/}
for short). According to \cite{MillerSTMCapacity} each individual
could remember $7 \pm 2$ objects in his or her STM at each moment
(a.k.a. Miller's law). Thus we assume the STM capacity to be 7
objects. Finally, we assume that $U$ can use paper and pen for those
cases where $U$ needs to keep track of more objects than can be fit
into STM.

User $U$ can clean the clusters output by the machine part in many
different ways. In this paper, based on what we have seen users
do in industry, we propose that $U$ clean in two stages. The first
stage splits the clusters recursively until all resulting clusters are
``pure'', i.e., each containing only the values of a single real-world
entity (though often not all such values). The second stage then
merges clusters that refer to the same entity.
\vspace{-2mm}\begin{example}
	Suppose the machine part produces clusters 1-2 in Figure
	\ref{humanex}. The split stage splits cluster 1 into clusters 3-4,
	cluster 2 into clusters 5-6, cluster 6 into clusters 7-8, then
	cluster 8 into clusters 9-10 (see the solid arrows). The output of
	the split stage is the set of pure clusters 3, 4, 5, 7, 9, 10. The
	merge stage then merges clusters 3 and 5 into cluster 11, and
	clusters 4 and 9 into cluster 12 (see the dotted arrows). The end
	result is the set of clean clusters 11, 12, 7, 10.
\end{example}

\vspace{-4mm}\subsection{The Split Stage}\label{secsplit}

\begin{figure}
	\centering
	\includegraphics[height=2.2in]{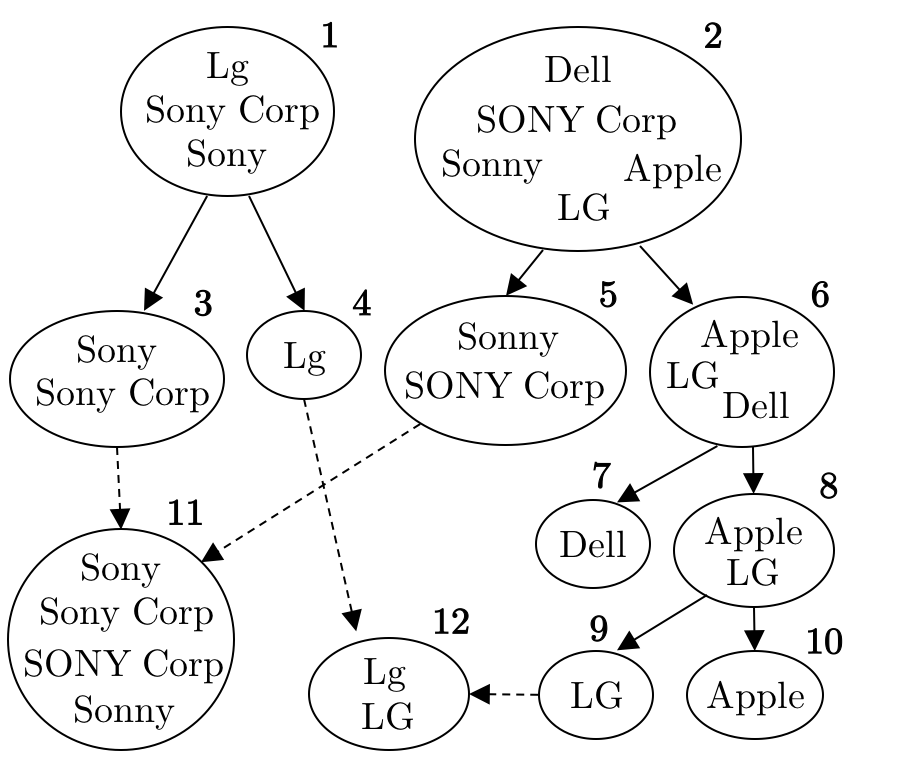}
	\vspace{-2mm}\caption{An illustration of the split and merge stages.}\label{humanex}
\end{figure}

We now describe the split stage (Section \ref{secmerge} describes the
merge stage). First, we define a dominating entity of a cluster $c$ to
be the one with the most values in $c$ (henceforth we use ``value''
and ``string'' interchangeably). Formally,
\vspace{-1mm}\begin{definition}[Dominating entity]\label{dom}
	Let $\mathcal{G}$ be a partition of cluster $c$ into groups of
	values $G_1, \ldots, G_n$, such that all values in each group refer
	to the same real-world entity and different groups refer to
	different entities. Then the dominating entity of $c$ is the entity
	of the group with the largest size: $G_k = arg\,max_{G_i\in
		\mathcal{G}}\ |G_i|$. Henceforth, we will use $dom(c)$ (or $e*$
	when there is no ambiguity) to denote the dominating entity of
	$c$.
\end{definition}
\vspace{-1mm}In Figure \ref{humanex}, {\em dom(Cluster 1)} and {\em dom(Cluster 2)} are
Sony Corporation.  Cluster 6 has three candidates; we break tie
by randomly selecting one to be the dominating entity. 

Let $C$ be the set of clusters output by the machine part. Our key
idea for the split stage is that if the machine part has been
reasonably accurate, then any cluster $c\in C$ is likely to be
dominated by $dom(c)$. If so, user $U$ can clean $c$ by moving all the
values in $c$ that {\em do not\/} refer to $dom(c)$ into a new cluster
$d$, then clean $d$, and so on.

\setlength{\algindent}{1em}
\begin{algorithm}[t]
	\begin{scriptsize}
		\caption{Split Phase} 
		\label{splitcode}
		\begin{flushleft}
			\textbf{Procedure} Split($C$)\\
			\textbf{Input:\,\,\,\,\,} a set of clusters $C=\{c_1, \dots, c_n\}$, output 
			by machine \\
			\textbf{Output:} a set of clean clusters $D = \{d_1, \dots, d_m\}$ s.t. 
			$\cup_{i} c_i = \cup_{j} d_j$
		\end{flushleft}
		\begin{algorithmic}[1]
			\Indstate {$D \leftarrow \emptyset$}
			\Indstate {\textbf{for each} cluster $c \in C$ \textbf{do} $D \leftarrow D 
				\cup \text{SplitCluster}(c)$}
			\Indstate {\textbf{return} $D$}
		\end{algorithmic}
		\-\hspace{-1cm}\newline
		\vspace*{-4mm}
		\begin{flushleft}
			\textbf{Procedure} SplitCluster($c$)\\
			\textbf{Input:\,\,\,\,\,} a cluster $c$\\
			\textbf{Output:} a set of clean clusters $G = \{g_1, \dots, g_p\}$ 
			$\cup_{k} g_k = c$
		\end{flushleft}
		\begin{algorithmic}[1]
			\Indstate {\textbf{if} $|c| = 1$ \textbf{then} \textbf{return} $\{c\}$}
			\Indstate {\uaispure($c$) // at the end, user selects yes/no button}
			\Indstate {\textbf{if} yes button is selected \textbf{then} \textbf{return} 
				$\{c\}$}
			\Indstate {\uafinddom($c$) // at the end, user knows $\underline{e^*}$ 
				and $\underline{\alpha}$}
			\Statex {\-\hspace{1.2em} // or ``mark values" button}
			\Indstate {\textbf{if} ``clean mixed cluster" button is selected 
				\textbf{then}}
			\Indstate[2] {\textbf{return} Merge($c$) // $\underline{\alpha} < 0.1$ in 
				this case}
			\Indstate {\underline{MarkValues}($c, \underline{e^*}, 
				\underline{\alpha}$)}
			\Statex {\-\hspace{1.2em} // at the end, user selects ``create/clean new 
				cluster"}
			\Statex {\-\hspace{1.2em} // or ``create new cluster / clean old cluster" 
				button}
			\Indstate {Move all marked values in $c$ into a new cluster $d$}
			\Indstate {\textbf{if} ``create/clean new cluster" button is selected 
				\textbf{then}}
			\Indstate[2] {\textbf{return} $c \cup \text{SplitCluster}(d)$ // 
				$\underline{\alpha} \geq 0.5$}
			\Indstate {\textbf{else} \textbf{return} $\text{SplitCluster}(c) \cup d$ // 
				$\underline{\alpha} < 0.5$}
		\end{algorithmic}
		\-\hspace{1cm}\newline
		\vspace*{-4mm}
		\begin{flushleft}
			\textbf{Procedure} \underline{MarkValues}($c, \underline{e^*}, 
			\underline{\alpha}$)\\
			\textbf{Input:\,\,\,\,\,} a cluster $c$, dominating entity $\underline{e^*}$ 
			and purity $\underline{\alpha}$ of $c$\\
			\textbf{Output:} a set of values in $c$ will be selected by the user
		\end{flushleft}
		\begin{algorithmic}[1]
			\Indstate {Let $L$ be the list of values in $c$, displayed on GUI}
			\Indstate {\textbf{if} $\underline{\alpha} \geq 0.5$ \textbf{then}} 
			\Indstate[2] {\textbf{for} $i \leftarrow 1, \dots, |c|$ \textbf{do}}
			\Indstate[3] {\uafocus($L[i]$), \textbf{if not} \uamatch($L[i], 
				\underline{e^*}$) 
				\textbf{then} \uaselect($L[i]$)}
			\Indstate {\textbf{else} \textbf{for} $i \leftarrow 1, \dots, |c|$ \textbf{do}}
			\Indstate[2] {\uafocus($L[i]$), \textbf{if} \uamatch($L[i], \underline{e^*}$) 
				\textbf{then} 
				\uaselect($L[i]$)}
		\end{algorithmic}
	\end{scriptsize}
\end{algorithm}

Specifically, for each cluster $c\in C$, user $U$ should (1) check if
$c$ is pure; if yes, stop; (2) otherwise find the dominating entity
$dom(c)$; (3) move all values in $c$ that {\em do not\/} refer to
$dom(c)$ into a new cluster $d$; then (4) apply Steps 1-3 to cluster
$d$ (cluster $c$ has become pure, so needs no further splitting).
This recursive procedure will split the original cluster $c$ into a
set of pure clusters. It is relatively easy for human users to
understand and follow, and as we will see in Section \ref{estimatecosts}, it is also
highly amenable to cost analysis.
\vspace{-1mm}\begin{example}
	Given cluster 1 in Figure \ref{humanex}, user $U$ splits it into the
	pure cluster 3, which contains only the values of the dominating
	entity Sony Corporation, and cluster 4, which contains all remaining
	values of cluster 1. A similar recursive splitting process applies
	to cluster 2. (Note that cluster 6 has three dominating-entity
	candidates, so we break tie randomly and select Dell to be the
	dominating entity.)
\end{example}
We now optimize the above procedure in three ways. First, if $c$ is a
singleton cluster, then we do not invoke the above splitting
procedure, because $c$ is already pure. Second, there are cases where
the number of values referring to $dom(c)$ is less than 50\% of
$|c|$. Formally, we define 
\vspace{-1mm}\begin{definition}[Cluster purity]\label{cpuritydef} The purity of a
	cluster $c$ is the fraction of the values in $c$ that refer to
	$dom(c)$. Henceforth we will use $p(c)$ (or $\alpha$ when there is
	no ambiguity) to denote the purity of $c$.
\end{definition}
\vspace{-1mm}For example, in Figure \ref{humanex}, the purity of cluster 2 is 2/5 =
0.4 $<$ 0.5. In such cases, instead of moving all values in $c$ that
do not refer to $dom(c)$, as discussed so far, it is less work for the
user to move the values that do refer to $dom(c)$ into a new cluster
$d$ (e.g., for cluster 2, $U$ should move ``Sonny'' and ``SONY Corp'',
instead of the other three values).

Finally, if $p(c)$ is below a threshold, currently set to 0.1, then
$c$ is very ``mixed'', with each entity having less than 10\% of the
values. In this case, we have found that instead of splitting $c$, it
is often more effective to apply the \merge\ procedure described in
Section \ref{secmerge} to $c$. This produces a set of pure clusters that are then
fed to the merge stage.

\paragraph{Basic User Operations:} To implement the above solution, we
define the following five basic user operations:

\vspace{2mm}\noindent\textbullet\ {\bf \underline{focus}(a)}: User $U$
moves his or her focus to a particular object $a$ on the GUI or on the
paper, such as a cluster, a value within a cluster, a GUI button, a
number on the paper, etc. Intuitively, user $U$ will shift his
or her attention from one object to another on the GUI or the paper,
and that incurs a certain amount of time.  This operation is designed
to capture this physical action (and its cost).

\vspace{2mm}\noindent\textbullet\ {\bf \underline{select}(a)}: User
$U$ selects an object $a$ on the GUI (e.g., a cluster, a value, a GUI
button, etc.) by moving the mouse pointer to that object and clicking
on it, or pressing a keyboard button (e.g., Page Up, Page Down). This
operation is designed to capture this physical action (and its cost). 

\vspace{2mm}\noindent\textbullet\ {\bf \underline{match}(x,y)}: Given
two values, or a value and a real-world entity (in $U$'s short-term
memory), $U$ determines if they refer to the same real-world entity.

\vspace{2mm}\noindent\textbullet\ {\bf \underline{isPure}(c)}: $U$
examines cluster $c$ to see if it is pure (i.e., if it is
clean). Specifically, we assume the values in $c$ is listed (e.g., on
the GUI) as a list of values $L$. User $U$ reads the first value of
$L$, maps it to an entity $e$, then scans the values in the rest of
$L$. As soon as $U$ sees a value that does not refer to $e$, the
cluster is not pure, $U$ stops and returns false. Otherwise $U$
exhausts $L$ and returns true.

\vspace{2mm}\noindent\textbullet\ {\bf \underline{findDom}(c)}: finds
the dominating entity $dom(c)$ and the purity $p(c)$ of a cluster
$c$. If $|c|\leq 7$, the size of the short-term memory (STM), then $U$
does this entirely in STM.  Specifically, $U$ scans the list of values
in $c$, maps each value into an entity, and keeps track of the number
of times $U$ has encountered a particular entity. Then $U$
returns the entity with the highest count $g$ as the dominating one,
and $g/|c|$ as the purity of cluster $c$. If $|c|>7$ then $U$ proceeds
as above, but uses paper and pen to keep track of the counts of the
encountered entities. 

\paragraph{The Split Procedure:} Algorithm \ref{splitcode} describes
\mysplit, a procedure that uses the above five operations to implement
the split stage. \mysplit\ takes the set of clusters output by the
machine part, then applies the \splitcluster\ procedure to each
cluster.  We distinguish two kinds of procedures: GUI-driven and
human-driven. \mysplit\ and \splitcluster\ are GUI-driven, i.e.,
executed by the computer. A GUI-driven procedure, e.g., \splitcluster,
may call human-driven procedures e.g., \underline{\sf isPure},
\underline{\sf findDom}, then pass control to user $U$ to execute
those procedures. To distinguish between the two, we underline the
names of human-driven procedures.

Algorithm \ref{splitcode} shows that \splitcluster\ handles the corner
case of singleton clusters (Step 1), then calls \underline{\sf isPure}
and asks user $U$ to take over (Step 2). At the end of this procedure, 
$U$ would have selected either ``yes'' or ``no'' button, indicating
whether the cluster is pure or not. In the former case, \splitcluster\ 
terminates, returning the pure cluster (Step 3). Otherwise, it calls
\underline{\sf findDom} (Step 4), and so on. Note that at the end
of \underline{\sf findDom}, user $U$ knows the dominating entity
$e*$ and the purity $\alpha$, but the computer does not know these. 
Hence these quantities (and all quantities that only $U$ know) are 
shown as underlined, e.g., $\underline{e*}$, $\underline{\alpha}$.

\vspace{-5.3mm}\subsection{The Merge Stage}\label{secmerge}
Given a set of pure clusters output by the split stage, the merge
state merges clusters that refer to the same entity. Clearly, from
each cluster we can select just a single representative value (say the
longest string), then merge those (if we know how to merge those, we
can easily merge the original clusters). For example, in Figure
\ref{humanex} the split stage produces clusters 3, 4, 5, 7, 9, and
10. To merge them, it is sufficient to consider merging the values
``Sony Corp'', ``Lg'', ``SONY Corp'', ``Dell'', ``LG'', and
``Apple''. Henceforth we will consider this simpler problem of merging
$n$ values $v_1, \ldots, v_n$.

\begin{figure}[t]
	\centering
	\includegraphics[height=1.3in]{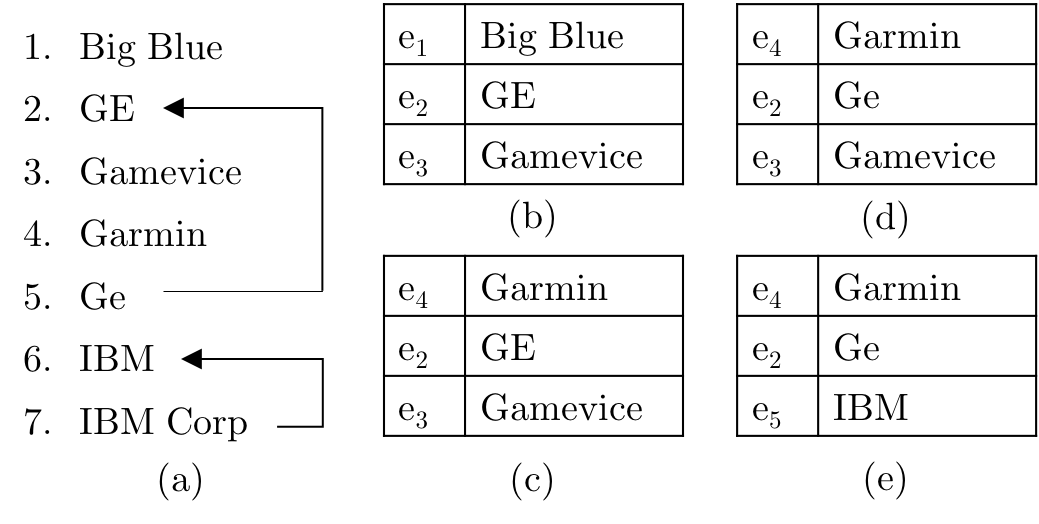}
	\vspace{-4mm}\caption{An example of local merging.}\label{lmerge}
	\vspace{-2mm}
\end{figure}

Naively merging by considering all pair takes quadratic
time. To address this problem, we propose a two-step process. First,
$U$ does one pass through the list of values to do a ``local merging''
that merges matching values that are near one another. This reduces
$n$. Then $U$ does ``global merging'' that considers all pairs (of the
remaining values). {\em Both steps will exploit the parallel processing
	capability of short-term memory (STM).} We now describe these two
steps.

\paragraph{Local Merging:} This step uses STM to merge matching values
that are near one another. Specifically, first the set of values is
sorted. Currently we use alphabetical sorting, because matching values
often share the first few characters (e.g., IBM, IBM Corp). Figure
\ref{lmerge}.a shows such a sorted list $L$ (ignoring the arrows for now).

Next, user $U$ processes the values in $L$ top down. For each value,
$U$ stores it and the associated entity in STM. For the sake of this
example, assume STM can only store three such pairs. Figure
\ref{lmerge}.b shows a full STM after $U$ has processed the first
three values of the list.  Then when processing the 4th value,
``Garmin'', $U$ needs to evict the oldest pair from STM to make space
for ``Garmin'' (see Figure \ref{lmerge}.c).

Then when processing the 5th value, ``Ge'', $U$ realizes that its
entity, $e_2$, is already in STM, associated with a previous value
``GE''. So $U$ links ``Ge'' with ``GE'', and replaces the value ``GE''
in STM with the new value ``Ge'' (see Figure \ref{lmerge}.d). Next,
``IBM'' will be stored in STM, displacing ``Gamevice'' (Figure
\ref{lmerge}.e), and so on. At the end, $U$ has linked together
certain matching values (see the arrows in Figure \ref{lmerge}.a).

Algorithm \ref{lmergefig} describes local merging, which uses
previously defined user operations \underline{focus}(a) and
\underline{select}(a) (see Section \ref{secsplit}), as well as the following new
user operation:

\vspace{1mm}\noindent\textbullet\ {\bf \underline{memorize}(v)}: $U$
maps the input value $v$ to an entity $e$, then memorizes, i.e., stores
the pair $(e,v)$ in STM. Specifically, if a pair $(e,t)$ is already in
STM, $U$ replaces it with $(e,v)$, then exits, returning $(e,t)$ (see
Line 2 in Algorithm \ref{lmergefig}).  Otherwise, $U$ adds the pair
$(e,v)$ to STM, ``kicking'' the oldest pair out of STM to make
space if necessary.

\setlength{\algindent}{0.8em}
\begin{algorithm}[t]
	\begin{scriptsize}
		\caption{Local Merging} 
		\label{lmergefig}
		\begin{flushleft}
			\textbf{Procedure} \underline{LocalMerge}($L$)\\
			\textbf{Input:\,\,\,\,\,} a list of values $L$ sorted alphabetically\\
			\textbf{Output:} links among certain values in $L$ that match
		\end{flushleft}
		\begin{algorithmic}[1]
			\Indstate {\textbf{for} $i \leftarrow 1, \dots, |L|$ \textbf{do}}
			\Indstate[2] {$(\underline{e}, \underline{t}) \leftarrow$ 
				\uamemorize($L[i]$)}
			\Indstate[2] {\textbf{if} $\underline{e}$ is not null \textbf{then}}
			\Statex {\-\hspace{1.8em} // $L[i]$ maps to $\underline{e}$, 
				$\underline{e}$ is already in STM and associated with $\underline{t}$}
			\Indstate[3] {\uaselect($L[i]$)}
			\Indstate[3] {\uafocus($L[j]$), \uaselect($L[j]$) // $L[j]$ is the previous 
				value $\underline{t}$}
			\Indstate[3] {\uafocus(link button), \uaselect(link button)}
			\Statex {\-\hspace{0.5em} // at the end, user selects ``done local 
				merging" button}
		\end{algorithmic}
	\end{scriptsize}
\end{algorithm}

\begin{figure}[b]
	\centering
	\includegraphics[height=1in]{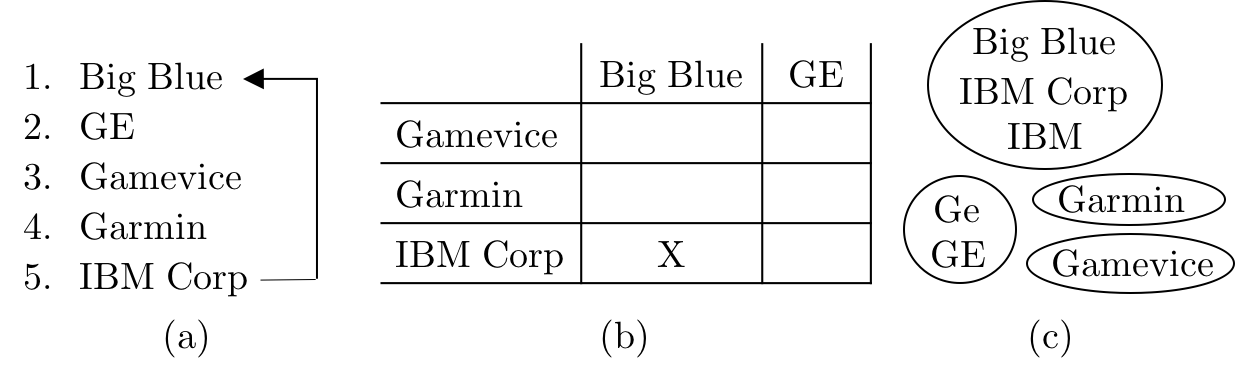}
	\vspace{-6mm}
	\caption{An example of global merging.}\label{gmerge}
\end{figure}

\paragraph{Global Merging:} After local merging, the original list of
values is consolidated, i.e., from each set of linked values we again
select just a single representative value (e.g., the longest
one). This produces a new shorter list, e.g., consolidating the list
in Figure \ref{lmerge}.a produces the shorter list in Figure
\ref{gmerge}.a (ignoring the arrow).

Let $L = [v_1, \ldots, v_n]$ be this new shorter list. Naively, user
$U$ can compare $v_1$ with $v_2, \ldots, v_n$, then $v_2$ with $v_3,
\ldots, v_n$, etc.  A better solution however is to exploit the
parallel processing capability of STM: read multiple values, say $v_1,
\ldots, v_k$, into STM all at once, then compare them all in parallel
with $v_{k+1}, \ldots, v_n$, etc.
\vspace{-1mm}\begin{example}
	Consider again the list in Figure \ref{gmerge}.a. User $U$ can read
	the first two values, ``Big Blue'' and ``GE'', into STM, then scan
	the rest of the values and match them with these two in parallel
	(using a GUI, see Figure \ref{gmerge}.b). If there is a match, e.g.,
	``IBM Corp'' and ``Big Blue'', then $U$ checks off the appropriate
	box (see Figure \ref{gmerge}.b).  At the end of the list, $U$ pushes
	a button to link the matching values.  Next, $U$ reads into STM the
	next two values, ``Gamevice'' and ``Garmin'', then match ``IBM
	Corp'' with these two. $U$ detects no more matches, thus wrapping up
	global merge. The system uses the results of both local and global
	merges to produce the final clusters shown in Figure \ref{gmerge}.c.
\end{example}
\vspace{-1mm}In practice, even though STM can hold 7 objects \cite{MillerSTMCapacity}, we found
that users prefer to read only 3 values at a time into STM. First, 7
values often take up too much horizontal space on the GUI (especially
if the strings are long), making it hard for users to
comprehend. Second, users want to reserve some STM capacity to read
and remember the values in the rows. 

As a result, we currently use $k=3$ in our global merge
procedure. Appendix \ref{asecplans} describes this procedure, which
uses user operations \underline{focus}(a), \underline{select}(a),
\underline{memorize}(v), as well as the following new user operation:

\vspace{1mm}\noindent\textbullet\ {\bf \underline{recall}(v)}: $U$
maps the input value $v$ to an entity $e$, then checks to see if $e$
is already in STM, returning $e$ and the associated value if yes,
and null otherwise. 

The \merge\ procedure (Appendix \ref{asecplans}) implements the entire
merge stage.  It calls \lmerge\ on the output of the split stage, then
\gmerge\ on the output of \lmerge. The following theorem (whose proof 
involves the verification sets of actions) shows the
correctness of the human part:
\vspace*{-2mm}\begin{theorem}\label{mytheory}
	Let $V$ be a set of strings to be normalized. Applying the \mysplit\ followed by \merge\
	procedures to any partition $\mathv$ of $V$ produces a set of clusters
	with 100\% precision and recall, assuming that the user correctly executes
	the operations, per their instructions. 
\end{theorem}

\vspace{-3mm}\section{Defining the Machine Part}\label{machine}

We now discuss the machine part of VN plans, which applies an
algorithm to cluster the input strings. Many algorithms can be used,
e.g., string clustering, string matching (SM), and entity matching
(EM). We first discuss using string clustering algorithms,
specifically HAC (hierarchical agglomerative clustering). Then we
discuss why existing SM/EM algorithms do not work well for our
purposes.

\vspace{-2mm}
\subsection{Using the HAC Clustering Algorithm}\label{usinghac}
\begin{figure}
	\vspace{-1mm}
	\centering\includegraphics[height=1.25in]{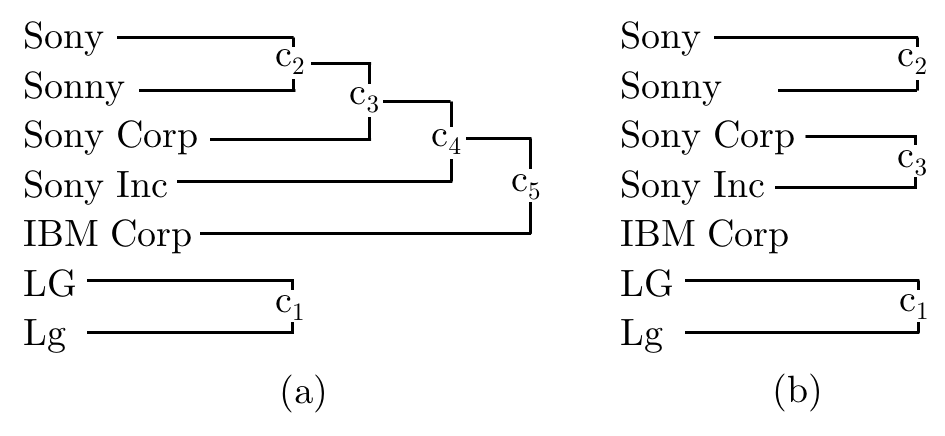}
	\vspace{-4mm}\caption{How HAC and HAC with a limit on cluster size
		work on the same dataset.}\label{hac}
	\vspace{-1mm}
\end{figure}

We consider using a generic clustering algorithm in the machine part. Many
such algorithms exist
\cite{clustersurvey,Xu:2005:SCA:2325810.2327433}. For now, we
consider hierarchical agglomerative clustering (HAC), because it is
easy to understand and debug, can achieve good accuracy, and commonly
used in practice. To cluster a set of values, HAC
initializes each value as a singleton cluster. Next, it finds the two
clusters with the highest similarity score (using a pre-specified
similarity measure), merges them, then repeats, until reaching a
stopping criterion, e.g., the highest similarity score falling below a
pre-specified threshold.
\vspace{-1mm}
\begin{example}\label{hacexam}
	Consider clustering the seven values in Figure \ref{hac}.a. HAC may
	first cluster ``LG'' and ``Lg'' into a cluster $c_1$, then ``Sony''
	and ``Sonny'' into $c_2$, then $c_2$ and ``Sony Corp'' into $c_3$,
	etc. The final result is clusters $c_1$ and $c_5$.
\end{example}
\vspace{-1mm}

\paragraph{Problems with Large Mixed Clusters Produced by HAC:} Using HAC ``as is'' however
does not work well, because it often produces large mixed clusters
that are time consuming for user $U$ to clean. Specifically, as
HAC iterates, it grows bigger clusters. Initially, when these clusters
are small, their quality is often quite good, because they often
group together syntactically similar values that refer to the same
entity (e.g., ``LG'' and ``Lg'', or ``Sony'' and ``Sonny'', see Figure
\ref{hac}.a).

As the clusters grow, however, they start attracting
``junk'', e.g., cluster $c_4$ attracts ``IBM Corp'' (Figure
\ref{hac}.a).  If the similarity measure used by HAC happens to be
``liberal'' for the data set at hand, HAC often grows large clusters
that are ``mixed'', i.e., containing the values of multiple entities.
It is very expensive for user $U$ to clean such clusters, using the
\mysplit\ and \merge\ procedure described in the previous section.

\begin{figure}[t]
	\vspace{-1mm}
	\centering
	\hspace*{-2mm}\includegraphics[width=0.85\columnwidth]{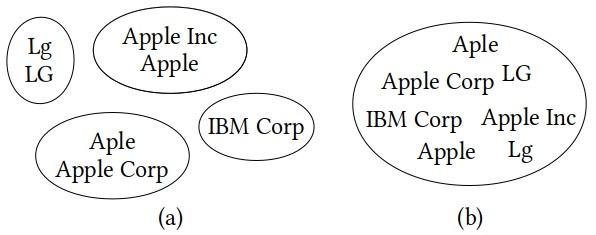}
	\vspace{-3mm}\caption{Cleaning (a) is less work than cleaning (b).}\label{morework}
	\vspace{-1mm}
\end{figure}

\paragraph{Proposed Solution:} Ideally, HAC should stop before its clusters
become too ``mixed''.  If HAC's clusters are smaller but pure (e.g.,
Figure \ref{morework}.a), then $U$ mostly just have to merge these
clusters using a few mouse clicks.  However, if HAC's clusters are larger but
``mixed'' (e.g., Figure \ref{morework}.b), then $U$ would need to
split them up into pure clusters, before merging them.  This incurs
far more mouse clicks and thus far more work.

Of course, we do not know when to stop HAC. To address this
problem, we introduce multiple HAC variations, each stopping at a
different time, then try to select a good one. Specifically, to
cluster $n$ values, we consider $n$ HAC variations, where the $i$-th
variation, denoted HAC(i), limits the cluster size to at most $i$. In
each iteration, HAC(i) finds the two clusters $c$ and $d$ with the
highest similarity score, then merges them if $|c\cup d|\leq i$.
Otherwise, HAC(i) finds the two clusters with the next highest score,
and merges them if the resulting size is at most $i$, and so on.
HAC(i) terminates when it cannot find any more clusters to merge.
\vspace{-1mm}\begin{example}
	Consider applying HAC(2) to the values in Figure \ref{hac}.a. HAC(2)
	first forms cluster $c_1$, then $c_2$, exactly as the normal
	HAC. Then normal HAC goes on to form cluster $c_3$ in Figure
	\ref{hac}.a, but HAC(2) cannot, because $c_3$'s size exceeds
	2. Instead, HAC(2) find the next two clusters with the highest
	similarity score. Suppose these are the singleton clusters for
	``Sony Corp'' and ``Sony Inc''. Then HAC(2) merges them to form
	cluster $c_3$ in Figure \ref{hac}.b. At this point HAC(2) cannot
	form any more cluster, because any resulting cluster size would
	exceed 2. So it stops, returning the clusters in Figure \ref{hac}.b
	as the output.
\end{example}
\vspace{-1mm}HAC(1) produces the smallest but cleanest clusters (as they are
singleton). As we increase $i$, HAC(i) tends to produce bigger but
less clean clusters. Typically, there exists an i* such that
HAC(i*)'s clusters are still so clean that they help user $U$, but
HAC(i*+1)'s clusters are already ``too dirty'' to help (e.g., $U$
would need to split them extensively before her or she can
merge). This roughly corresponds to the point where we want HAC to
stop.  HAC(i*) thus is the ``best'' HAC variation for
the current data set.

To find HAC(i*), we pair HAC(1), $\ldots$, HAC(n) with \mysplit\ and
\merge\ to form $n$ end-to-end plans. Sections \ref{estimatecosts} and 
\ref{searching} show how
to estimate the costs of these plans and find the one with the least
estimated cost.

\vspace{-2mm}\subsection{Limitations of SM/EM Solutions}\label{limsmem}\label{limitsol}
We are now in a position to explain why existing string matching (SM)
and entity matching (EM) solutions do not work well in our context.
(Section \ref{eval} shows experimentally that \winston\ with HAC
outperforms these solutions.)

At the core, VN is an SM problem. So SM solutions can be used
in the machine part. EM solutions can also be used, by
limiting each entity to be a string. Many such solutions have been
developed, e.g., \transer\ \cite{wangtransitive2013}, \magellan\ \cite{magellan}, 
\falcon\ \cite{falcon-sigmod17}, \waldo\ \cite{waldo} (see
Section \ref{relatedwork}).

These SM/EM solutions (e.g., \magellan, \falcon) typically output a
set of matches. One way to use them is to ask user $U$ to verify
certain matches, then infer even more matches using match
transitivity. For example, given 5 string $a, b, c, d, e$,
suppose a solution outputs $a = b$ and $a = c$ as matches. If $U$ has
verified these matches, then we can infer $b = c$ as another match. A
recent work, \transer\ \cite{wangtransitive2013}, exemplifies this approach.  {\em A
	serious problem, however, is that we cannot guarantee 100\% recall,
	as shown experimentally in Section \ref{eval}} For example, no user
verification and match transitivity on the outputs $a = b, a = c$ can
help us infer $d = e$ (assuming this is also a correct match). Thus,
these solutions are not appropriate for \winston.

Another way to use existing SM/EM solutions is to cluster the input
strings in a way that respect the output matches. The work \cite{mcl}
describes multiple ways to do this. Continuing with
the above example, given the output matches $a = b, a = c$, we can
cluster the five input strings into, say, 3 clusters $\{a,b,c\},
\{d\}, \{e\}$. User $U$ can verify/clean these clusters, as discussed
in the human part. {\em A serious problem here, however, is that this
	approach often produces large mixed clusters, which are very time
	consuming for $U$ to clean, as shown experimentally in Section \ref{eval}.}
With HAC, we solve this problem by modifying HAC to stop early to
produce clean clusters (see Section \ref{usinghac}). But there is no obvious way
to modify the clustering algorithms in \cite{mcl} to stop early such that
they produce relatively clean clusters and the quality of these clusters
can be estimated (e.g., see Section \ref{estimatecosts}). 

The above works provide no GUI, or very basic inefficient GUIs for
user feedback, e.g., \falcon\ and \transer\ ask users to label string
pairs as match/non-match.  A recent work, \waldo\ \cite{waldo}, considers
a far more efficient GUI, which displays 6 strings so that a user can
cluster all of them in one shot. As such, its ``human'' part is more
similar to ours. But its ``machine'' part considers a very different
optimization problem: namely minimizing crowdsourcing cost (e.g.,
clustering 6 strings incurs the same monetary cost, regardless of
which human user does it). Thus, it cannot be used in \winston, which
focuses on minimizing the effort of human users.

\vspace{-3mm}\section{Estimating Plan Costs}\label{estimatecosts}
We now discuss estimating the cost of a plan, which is the total time
user $U$ spends in the human part to clean the clusters output by the
machine part. As we will see below, the key idea is to estimate the
quality of these clusters, then use that to estimate the time needed
to clean them.

Specifically, let $V = \{v_1, \ldots, v_n\}$ be the set of input
values, and $p_1, \ldots, p_n$ be the plans that we will consider, where
each plan $p_{\lambda}$ applies HAC($\lambda$) to $V$ to obtain a set
of clusters $C_{\lambda}$, then employs a user $U$ to clean
$C_{\lambda}$, using \mysplit\ and \merge. Let
$\clam = \{c_1, \ldots, c_{\nlam}\}$. Then the cost of $p_{\lambda}$ (i.e.,
the time for $U$ to clean $\clam$) can be expressed as
$cost(\plam)  =  \sum_{i=1}^{\nlam} time(SplitCluster(c_i))\ +\ 
time(LocalMerge(L))\ +\\
time(GlobalMerge(T)),$ where $L$ is a list of values summarizing the output 
of {\sf SplitCluster}, and $T$ is a list summarizing the output of {\sf
	LocalMerge}. We now estimate these quantities.

\paragraph{Estimating the Cost of SplitCluster:} We need to estimate
$SplitCluster(c_i)$ for each cluster $c_i\in \clam$. To do this, we
make two assumptions:

\vspace{1mm}\noindent(1) All clusters $c_1, \ldots, c_{\nlam}$ produced by 
$p_{\lambda}$ have 
the same cluster purity $\alpha_{\lambda}$ (which is defined in Definition 
\ref{cpuritydef}).\vspace{1mm}\\
\noindent(2) When we use {\sf SplitCluster} to split a
cluster $c_i$ (produced by $p_{\lambda}$) into a pure cluster containing all
values of the dominating entity and a ``mixed'' cluster containing all
the remaining values, the ``mixed'' cluster also has purity
$\alpha_{\lambda}$. When we split this ``mixed'' cluster, the
resulting ``mixed'' cluster also has purity $\alpha_{\lambda}$, and so
on.

These are obviously simplifying assumptions. However, they reflect the
intuition that each plan \hlam\ produces
clusters of a certain quality level, and that this quality level can
be captured by a single number, $\alphalam$, which is the purity of
all the clusters. Further, they allow us to efficiently estimate plan
costs. Finally, Section \ref{eval} empirically shows that with these
assumptions we can already find good plans.

Next, we use the above assumptions to estimate the cost of $SplitCluster(c_i)$. 
Suppose that we already know $\alphalam$ (we show
how to estimate $\alphalam$ later in this section), and that
$\alphalam \geq 0.5$, then when splitting $c_i$ using {\sf
	SplitCluster}, user $U$ creates two clusters: a pure dominating
cluster $c_{i,w}$ of size $\alpha_{\lambda} \psi_i$, where $\psi_i$ is
the size of $c_i$, and a remainder cluster $c_{i,u}$ of size
$(1-\alpha_{\lambda}) \psi_i$. If $|c_{i,u}| > 1$, we assume that its
purity is also $\alpha_{\lambda}$. $U$ then splits $c_{i,u}$,
etc. After $\beta_i$ splits, $U$ has created $\beta_i +1$ clusters of
sizes $\alpha_{\lambda} \psi_i, \alpha_{\lambda} (1 -
\alpha_{\lambda}) \psi_i , \dots, \alpha_{\lambda} (1 -
\alpha_{\lambda})^{\beta_i - 1} \psi_i, (1 -
\alpha_{\lambda})^{\beta_i} \psi_i$, such that the last cluster has a
single element. Thus we can estimate $\beta_i$ as $\lfloor
-\log^{\psi_i}_{1 - \alpha_{\lambda}} \rfloor$.  Since we can split a
cluster of size $\psi_i$ at most $\psi_i - 1$ times, we set $\beta_i =
\min(\psi_i - 1, \lfloor -\log^{\psi_i}_{1 - \alpha_{\lambda}}
\rfloor)$.

Recall that Section \ref{secplans} defines seven user operations:
\underline{focus}(a), \underline{select}(a), \underline{match}(x,y),
\underline{isPure}(c), \underline{findDom}(c),
\underline{memorize}(v), and \underline{recall}(v) (see Table \ref{paratable}). Let 
$\rof, \ros,
\rom, \rop, \rod, \roz$, and $\ror$ be their costs (i.e., times),
respectively. As we will see later, the cost $\rop$ of
\underline{isPure}(c) is a function of $\psi_i$, the size of $c$, and
$\alphalam$, the purity of $c$. Hence, abusing notations, we will
denote this cost as $\rop(\psi_i,\alphalam)$ for cluster
$c_i$. Similarly, the cost $\rod$ of \underline{findDom}(c) is a
function of the size of $c$, and will be denoted as $\rod(\psi_i)$ for
cluster $c_i$. The remaining five costs (e.g., $\rof, \ros$, etc.)
will be constants. Now we can estimate the cost of
$SplitCluster(c_i)$ where $\alpha_\lambda \geq 0.5$ as
\vspace{-2mm}\begin{equation*}
	\scriptsize
	\begin{aligned}
		&cost_{SplitCluster}(c_i; \alpha_\lambda \geq 0.5)= 
		\textstyle\sum_{j=1}^{\beta_i} [\rho_p\big( (1 - \alpha_{\lambda})^{j-1} 
		\psi_i ,\alpha_{\lambda}\big) +\rho_f \\
		&  + \rho_s +\rho_d\big( (1 - \alpha_{\lambda})^{j-1} \psi_i\big) + 
		\rho_f + \rho_s + ( (1 - \alpha_{\lambda})^{j-1} \psi_i )  (\rho_f +\rho_m \\
		&+ (1 - \alpha_{\lambda}) \rho_s)+ \rho_f + \rho_s ]. \\
	\end{aligned}
\end{equation*}
Appendix \ref{aestimatecosts} discusses deriving the above formula, and 
computing the
cost for the cases $\alpha_\lambda \in [0.1,0.5)$ and $\alpha_\lambda
< 0.1$.

\newcommand{\fcal}{l}
\begin{table}
	\setlength\tabcolsep{1.5pt}
	\scriptsize
	\centering
	\hspace*{-5mm}
	\begin{tabular}{|c|c|c|c|} \hline
		\textbf{Parameter} & \textbf{Meaning} & \textbf{Model} & 
		\begin{tabular}{@{}c@{}} \textbf{Estimation}\\\textbf{Method} 
		\end{tabular}\\ \hline
		$\alphalam$ & \multicolumn{1}{|\fcal|}{\begin{tabular}{@{}l@{}} Cluster 
				purity for HAC($\lambda$) \end{tabular}}& $\alphalam = a 
		\lambda^b$ & 
		User feedback\\ \hline
		$\rho_f$ & \multicolumn{1}{|\fcal|}{Cost of 
			\underline{focus}(a)} & A 
		constant & Set to 0.5 \\ 
		$\rho_s$ & \multicolumn{1}{|\fcal|}{Cost of \underline{select}(a)} & A 
		constant & Set to 0.5 \\ 
		$\rho_m$ & \multicolumn{1}{|\fcal|}{Cost of \underline{match}(x,y)} & A 
		constant & User feedback \\ 
		$\rho_p$ & \multicolumn{1}{|\fcal|}{Cost of \underline{isPure}(c)} & 
		$\rho_p(\psi, \alpha) = \gamma \psi \alpha + \gamma_0$& User feedback 
		\\ 
		$\rho_d$ & \multicolumn{1}{|\fcal|}{Cost of \underline{findDom}(c)} & 
		\begin{tabular}{@{}c@{}}$\rho_d(\psi) = \eta_1 \psi \text{ if }\psi \leq 	
			|STM|,$\\$\eta_2 \psi^2 + \eta_3$ o.w. \end{tabular} & User feedback \\ 
		$\rho_z$ & \multicolumn{1}{|\fcal|}{Cost of \underline{memorize}(v)} & A 
		constant & Set to 0.4 \\ 
		$\rho_r$ & \multicolumn{1}{|\fcal|}{Cost of \underline{recall}(v)} & A 
		constant & Set to $\rho_z$ \\ \hline 
		$\tau$ & \multicolumn{1}{|\fcal|}{\begin{tabular}{@{}l@{}} Shrinkage factor 
				for local merge\end{tabular}} & A constant & Set to 0.98 \\ 
		$\xi$ & \multicolumn{1}{|\fcal|}{\begin{tabular}{@{}l@{}} Hit factor for 
				global merge\end{tabular}} & A constant & Set to 0.1 \\ \hline
	\end{tabular}
	\vspace{0.5mm}\caption{Parameters for our cost models.}\label{paratable}
	\vspace{-5mm}
\end{table}

\paragraph{Estimating the Cost of LocalMerge:} Recall that
HAC($\lambda$) produces the set of clusters $\clam = \{c_1, \ldots,
c_{\nlam}\}$, and that $\beta_i$ is the total number of splits user
$U$ performs in {\sf SplitCluster} for each cluster $c_i$.  Then at
the end of the split phase, $U$ has produced a set of $\rlam$ pure
clusters, where $r_{\lambda}=\sum_{i=1}^{n_{\lambda}} (\beta_i +
1)$. Assuming that executing {\sf LocalMerge} on any list will shrink
its size by a factor of $\tau$ (currently set to 0.98), we can
estimate the time of executing {\sf LocalMerge} on the output of the
split phase as $r_\lambda \rho_z + r_\lambda (1 - \tau) (3\rho_f +
2\rho_s) +\rho_f + \rho_s$ (see Appendix \ref{aestimatecosts} for an 
explanation).

\paragraph{Estimating the Cost of GlobalMerge:} {\sf LocalMerge}
produces $r'_\lambda = \tau r_\lambda$ pure clusters to which user $U$
will apply {\sf GlobalMerge}. Recall that {\sf GlobalMerge} takes the
first three values in the input list $L$, displays them in three
columns, then asks $U$ to go through the rest of the values of $L$ and
check a box if any value matches the values of the columns (see Figure
\ref{gmerge}), and so on.  We assume that in each such iteration, for
each column, $\xi$ values will match, resulting in $\xi$ checkboxes
being marked. Then we can estimate the cost of {\sf GlobalMege} as
$\sum_{j=1}^{\left\lfloor 1 / (3\xi) \right\rfloor} [ 3 \rho_z +
(r'_\lambda - 3 (j - 1) \xi r'_\lambda - 3)\rho_r + 3 (\xi r'_\lambda
- 1)(\rho_f + \rho_s) + \rho_f + \rho_s ]$ (see Appendix
\ref{aestimatecosts}).

\paragraph{Estimating the Cluster Purity $\alphalam$:} Recall that we
assume all clusters $c_1, \ldots, c_n$ produced by \hlam\ have the
same cluster purity $\alphalam$. Using set-aside datasets, we found
that $\alphalam$ could be estimated reasonably well using a power-law
function $a\lambda^b$ (where $b$ is negative, see Table \ref{paratable}). To 
estimate $a$ and
$b$, we compute $\lambda_{10}$ and $\lambda_{20}$. To compute
$\lambda_{10}$, we apply HAC(10) to the set of input values to obtain
a set $C_{10}$ of clusters. Next, we randomly sample 3 clusters of size 10
from $C_{10}$ (if there are less than 3 such clusters, we
select the three largest). Next, we show each cluster to user $U$, ask him/her
to identify all values referring to the dominating entity, then use
those to compute the cluster purity. Finally, we take the average
purity of these clusters to be $C_{10}$. We proceed similarly to
compute $C_{20}$.

We now have three data points: (1,1), (10, $\lambda_{10}$), and
(20,$\lambda_{20}$), which we can use to estimate $a$ and $b$ in the
function $a\lambda^b$, using the ordinary least-squares method.

\paragraph{Estimating the Costs of User Operations:} Finally, we
estimate the costs of the seven user operations (see Table \ref{paratable}). The 
costs of \underline{focus}(a) and
\underline{select}(a) measure the times user $U$ focuses on an object
$a$ then selects it (e.g., by clicking a mouse button).  After a
number of timing with various users, we found that these times are
roughly the same for most users, and we set them to be $\rho_f =
\rho_s = 0.5$ seconds.  Similarly, we found the times of
\underline{memorize}(v) and \underline{recall}(v) to be roughly
constant, at $\rho_z = \rho_r = 0.4$ seconds respectively
(see Table \ref{paratable}).

The time $\rho_m$ of \underline{match}(x,y), however, while largely
not dependent on $x$ and $y$, does vary depending on user
$U$. Further, estimating the time $\rho_p$ of \underline{isPure}(c)
and time $\rho_d$ of \underline{findDom}(c) is significantly more
involved. To determine whether a cluster $c$ is pure, user $U$ needs
to examine at most $\alpha\psi$ values in $c$ (where $\psi$ is the
size of $c$) before he/she sees the first value not referring to
$dom(c)$. Hence, we model the time of \underline{isPure}(c) as
$\rho_p(\psi, \alpha) = \gamma \, \alpha \, \psi + \gamma_0$.

To find the dominating entity $e*$ of cluster $c$, we distinguish two
cases. If $\psi \leq |STM|$, then user $U$ can execute
\underline{findDom}(c) entirely in $U$'s short-term memory. In this
case the time is proportional to $\psi$. Otherwise $U$ needs to use
paper and pen, and we found that the time roughly correlates to
$\psi^2$. Thus, we model the time $\rho_d(\psi)$ of
\underline{findDom}(c) as $\eta_1 \psi$ if $\psi\leq |STM|$ and as
$\eta_2 \psi^2 + \eta_3$ otherwise.

All that is left is to estimate the cost $\rho_m$ of
\underline{match}(x,y), and the parameters $\gamma, \gamma_0, \eta_1,
\eta_2, \eta_3$ of the cost models of \underline{isPure}(c) and
\underline{findDom}(c). To do so, when running HAC(20) (to estimate
cluster purity $\alphalam$), we also ask user $U$ to perform a few
\underline{match}, \underline{isPure}, and \underline{findDom}
operations, then use the recorded times to estimate the above
quantities (see Appendix \ref{aestimatecosts}). Altogether, the time it takes for 
users
to calibrate cluster purity $\alphalam$ and the cost models of user
operations was mere minutes in our experiments (and was included in
the total time of our solution).

\vspace*{-1mm}
\section{Searching for the Best Plan}\label{searching}

Recall that to cluster the values $V = \{v_1, \ldots,
v_n\}$, we consider $n$ plan $p_1, \ldots, p_n$, where each plan
$p_{\lambda}$ applies HAC($\lambda$) to $V$ to obtain a set of
clusters $C_{\lambda}$, then employs a user $U$ to clean
$C_{\lambda}$.  We now discuss how to efficiently find the plan
$p_{\lambda*}$ with the least estimated cost.

Naively, we can (1) execute \hlam\ for each plan
$\plam$ to obtain $C_{\lambda}$, (2) apply the cost estimation
procedures in the previous section to $C_{\lambda}$ to
compute the cost of $\plam$, then (3) return the plan with the lowest
cost. Steps 2-3 take negligible times. Step 1 however applies HAC(1),
$\cdots$, HAC(n) separately to $V$, which altogether can take
a lot of time, e.g., 7.3 minutes for $|V| = 480$ and 1.1 hours for $|V|
= 960$ in our experiments.

To address this problem, we have developed a solution to jointly
execute HAC(1), $\cdots$, HAC(n), such that executing a plan can reuse
the intermediate results of executing a previous plan. Specifically,
we first execute HAC(n), i.e., the regular HAC. Recall that each
iteration $Iter_i$ of HAC(n) merges two clusters. Let $s(i)$ be the
size of the largest cluster at the end of $Iter_i$. Suppose there is a
$k$ such that $s(i)\leq \lambda$ but $s(i+1) > \lambda$. Then we know
that \hlam\ can reuse everything HAC(n) has produced up to $Iter_i$,
but cannot proceed to $Iter_{i+1}$.  So at the end of $Iter_i$ we save
certain information for \hlam\ (e.g., the merge commands so far, the
value $\lambda$), then continue with HAC(n). Once HAC(n) is done, we
go back to each saved point $\lambda$ and resume \hlam\ from there.
This strategy enables great reuse, especially for high values of
$\lambda$, e.g., slashing the time for 960 values from 1.1 hours to 18
secs.

\paragraph{Putting It All Together:} We can now describe the entire \cc\ system,
as used by a single user $U$. Given a
set of values $V$ to normalize, (1) \cc\ first calibrate the cluster
purity $\alphalam$ and the cost models. To do so, it runs HAC(10) and
HAC(20), asks user $U$ to perform a few basic tasks on sample clusters
from these algorithms, then use $U$'s results to calibrate (see
Section \ref{estimatecosts}). (2) \cc\ runs the above search
procedure to find a plan $p_{\lambda*}$ with the least estimated
cost. (3) Finally, \cc\ sends the output clusters of $p_{\lambda*}$ to
user $U$ to clean, using procedures \mysplit\ and \merge.\\
\vspace*{-5mm}
\section{Working with Multiple Users}\label{cwinstonoverview}
So far we have discussed how \winston\ works with a single user. In
practice, however, multiple users (e.g., people in the same team)
are often willing to jointly perform VN. We now discuss how to
extend \winston\ to divide the work among such users, to speed
up VN.

Consider the case of 3 users. Naively, we can divide the set of input
strings into 3 equal parts, ask each user to apply \winston\ to
perform VN for a part, then combine the three outputs to form a set of
clusters. We can obtain a canonical string from each cluster,
producing a new list of strings. Then we can divide this new list
among 3 users, repeat the process, and so on. This naive solution
however does not work well, because it often spreads matching strings,
i.e., those belonging to a golden cluster, among all 3 users, causing
much additional work in matching across the individual lists, in later
steps.

Intuitively, strings within a golden cluster should be assigned to a
single user, as much as possible. We have extended \winston\ to
realize this intuition. In the extension, \winston\ first briefly
interacts with each user to learn his/her profiles. Next, it uses
these profiles to search a plan space to find a good VN plan. Next, it
executes the machine part of this plan to produce a set $C$ of
clusters. It then divides $C$ among the users, such that each will
have roughly the same workload. The intuition here is that a cluster
in $C$ captures many strings that belong to the same golden cluster,
and is assigned to a single user. Next, \winston\ asks each
user to use \mysplit\ and \merge\ to clean the assigned clusters.
Finally, it obtains the set of (cleaned) clusters from all users, then
repeatedly performs a distributed version of \gmerge\ until all
clusters have been verified and cleaned. Appendix \ref{cwinston}
describes the algorithm in detail, provides the pseudo code,
and discusses cost estimation procedures for this version of \winston.\vspace*{-5mm}\\
\section{Empirical Evaluation}\label{eval}
We now evaluate \winston. Among others, we show that
\winston\ can significantly outperform existing solutions, that it can
leverage multiple users to drastically cut VN time, and that it can
scale to large datasets.

\begin{table}
	\setlength\tabcolsep{1.5pt}
	\scriptsize
	\centering
	\begin{tabular}{|l|c|l|l|} \hline
		\textbf{Name} & \textbf{Size} & \textbf{Description} & 
		\textbf{Sample Values} \\ \hline
		Nickname & 5132 & \begin{tabular}{@{}l@{}}Nicknames and \\some of their 
			typos\end{tabular} & \begin{tabular}{@{}l@{}} ``Cissy'', ``Fanny'', 
			``Frannie'' \end{tabular} \\ \hline
		Citation & 3000 & \begin{tabular}{@{}l@{}} Article citations from \\Google 
			Scholar and DBLP \end{tabular}& \begin{tabular}{@{}l@{}} ``caching 
			technologies for web \\applications c mohan vldb 2001''\end{tabular} \\ 
		\hline
		Life Stage & 199 & \begin{tabular}{@{}l@{}} Target life stage(s) \\of 
			products\end{tabular} & \begin{tabular}{@{}l@{}} ``Maternity'', 
			``Mothers'', 
			``Youth\textbar \\Young Professionals''\end{tabular} \\ \hline
		Big Ten & 74 & \begin{tabular}{@{}l@{}}Names of Big Ten \\Conference 
			colleges \end{tabular} & \begin{tabular}{@{}l@{}} ``University of Iowa'', 
			``UIowa'', \\``UM Twin Cities'' \\ \end{tabular} \\ \hline
	\end{tabular}
	\vspace{1mm}
	\caption{Datasets for our experiments.}\label{data}
	\vspace*{-5.5mm}
\end{table}
\vspace*{-2.5mm}
\subsection{Existing Manual/Clustering Solutions}
We first compare \winston\ with state-of-the-art manual and clustering
solutions (Section \ref{smemsolnlimit} considers string/entity matching solutions). 
We use the four datasets in Table
\ref{data}, obtained online and from VN tasks at a company. For each
dataset we manually created all correct clusters, to serve as the
ground truth. (We consider larger datasets later in Section \ref{addexpers}.)

\paragraph{The Existing Solutions:} We consider four solutions: \manual, \merge, \quack, and
\orefine. \manual\ is the typical manual method that we have observed
in industry. It can be viewed as performing the \gmerge\ method
(Section \ref{secmerge}). \merge\ is our own manual VN method, which
performs \lmerge\ then \gmerge.

\quack\ is a string clustering tool used extensively for VN at a
company. It also uses HAC like \winston, but does not place a limit on
the cluster size. We extended \quack\ by asking the user to clean the
clusters using \mysplit\ and \merge.  \merge\ and \quack\ can be
viewed as the two plans HAC(1) and HAC($n$) in the plan space explored
by \cc\ (where $n$ is the number of values to be normalized).

\orefine\ is a popular open-source tool to wrangle data
\cite{openrefine}. It uses several string clustering algorithms to
perform VN \cite{orclustering}. Among these, the most effective one
appears to be KNN-based clustering \cite{orclustering}. We extend this
algorithm to work with \mysplit\ and \merge\ (because the GUI provided
by \orefine\ is very limited).

\paragraph{Results:} Table \ref{exper1} shows the times of \winston\ vs.
the above four methods (in minutes), using a single user. For each
method we measure the total time the user spends cleaning the clusters
(for \clang\ this includes the calibration time). It is difficult to
recruit a large number of real users for these experiments, because
cleaning some datasets (e.g., Nickname) would take a few working
days. So we use synthetic users and each data point here is averaged
over 100 such users, see Appendix \ref{aeval}. (We use real users to
``sanity check'' these results in Section \ref{addexpers}.)

The table shows that \manual\ performs worst, incurring 3-6800
minutes. \merge\ performs much better,
especially on the two large datasets, incurring 4-1961 minutes,
suggesting that performing a local merge before a global merge is
important. \merge\ is clearly the manual method to beat.

\quack\ is a bit faster than \merge\ on Nickname (1808
vs. 1961), but slower on the remaining three datasets. \orefine's
performance is very uneven. It is a bit faster than \merge\
on Citation, but far slower on the other three datasets. 

In contrast, \winston\ performs much better than \merge. On Nickname
it saves 7.5 hours of user time (see the last column).
On Citation it saves 3.34 hours of user time. On Life Stage it is
comparable to \merge, and on Big Ten it is only 3 mins worse (due to
the overhead of user calibration time).

\clang\ also outperforms both \quack\ and \orefine. Importantly, in
all cases where \quack\ or \orefine\ performs worse than \merge,
\clang\ is able to select a good plan which allows it to outperform
\merge.

\begin{table} \setlength\tabcolsep{3pt} 
	\fontsize{7}{8}\selectfont
	\centering
	\hspace*{-1mm}
	\begin{tabular}{|l|c|c|c|c|c|c|c|} \hline
		\textbf{Dataset} & \textbf{Manual} & \textbf{Merge} &
		\textbf{Quack}& \textbf{OpenRefine} & \textbf{Winston} & \textbf{Savings} \\ \hline
		\textbf{Nickname} & 6800 & 1961 & 1808& $>$10000& 1512 & 7.5hrs\\ \hline
		\textbf{Citation} & 6585 & 1313 & 1371& 1280 & 1112 & 3.4hrs\\ \hline
		\textbf{Life Stage} & 13& 9& 15 & 12 & 9 & 0hrs\\ \hline
		\textbf{Big Ten}    & 3& 4 & 9& 22 & 7 & -3min
		\\ \hline
	\end{tabular}
	\vspace{1mm}
	\caption{Winston vs four existing solutions.} \label{exper1}
	\vspace{-9mm}
\end{table}

\begin{table}
	\begin{footnotesize}
		\setlength\tabcolsep{2pt}
		\centering
		\begin{tabular}{|l|c|c|c|c|c|} \hline
			\begin{tabular}{@{}l@{}}\textbf{Dataset} \end{tabular}& \textbf{1 User} & \textbf{3 Users} & 
			\textbf{5 Users} & \textbf{7 Users} & \textbf{9 Users} \\ \hline
			Nickname   & 1512	& 890 & 610 & 460 	& 412 	\\ \hline
			Citation 	   & 1112	& 428 & 278 & 212 	& 177 	\\ \hline
			Life Stage	 &  9		 & 6 	 & 6 	 & 6 	   & 6		 \\ \hline
			Big Ten		  &  7		  & 4 	  & 4 	  & 4 		& 4		  \\ \hline
		\end{tabular}
	\end{footnotesize}
	\vspace{1mm}
	\caption{The times of Winston with multiple users.}\label{winstonresults}
	\vspace{-5mm}
\end{table}

\vspace{-2mm}\subsection{Working with Multiple Users}

We have shown that \winston\ outperforms existing manual and
clustering methods. We now examine how \winston\ can leverage multiple
users to reduce VN time. Table \ref{winstonresults} shows that
\winston\ can leverage multiple users to drastically cut the VN time,
e.g., from 1512 minutes with 1 user to 412 with 9 users for Nickname, and from
1112 to 177 for Citation. The most significant reduction is achieved
early, e.g., from 1 to 3-5 users. After that, adding more users still
helps reduce the VN time, but only in a ``diminishing-return''
fashion.

\vspace{-2mm}\subsection{Limitations of SM/EM Solutions}\label{smemsolnlimit}
We now compare \winston\ to existing string matching (SM) and entity
matching (EM) solutions, specifically with \transer\ \cite{wangtransitive2013},
\falcon\ \cite{falcon-sigmod17}, and \magellan\ \cite{magellan}.

\paragraph{Comparing with TransER:} As discussed in Section \ref{limsmem}, 
there are two main ways to use SM/EM
solutions in our context. First, a solution can produce a set of
matches $M$, employ a user $U$ to verify certain matches in $M$, then
use match transitivity to infer even more matches. The work \cite{wangtransitive2013}
describes such a solution, which we call \transer.

\begin{figure}[t]
	\centering
	\vspace*{-2mm}
	\hspace*{-2mm}
	\includegraphics[width=0.9\columnwidth]{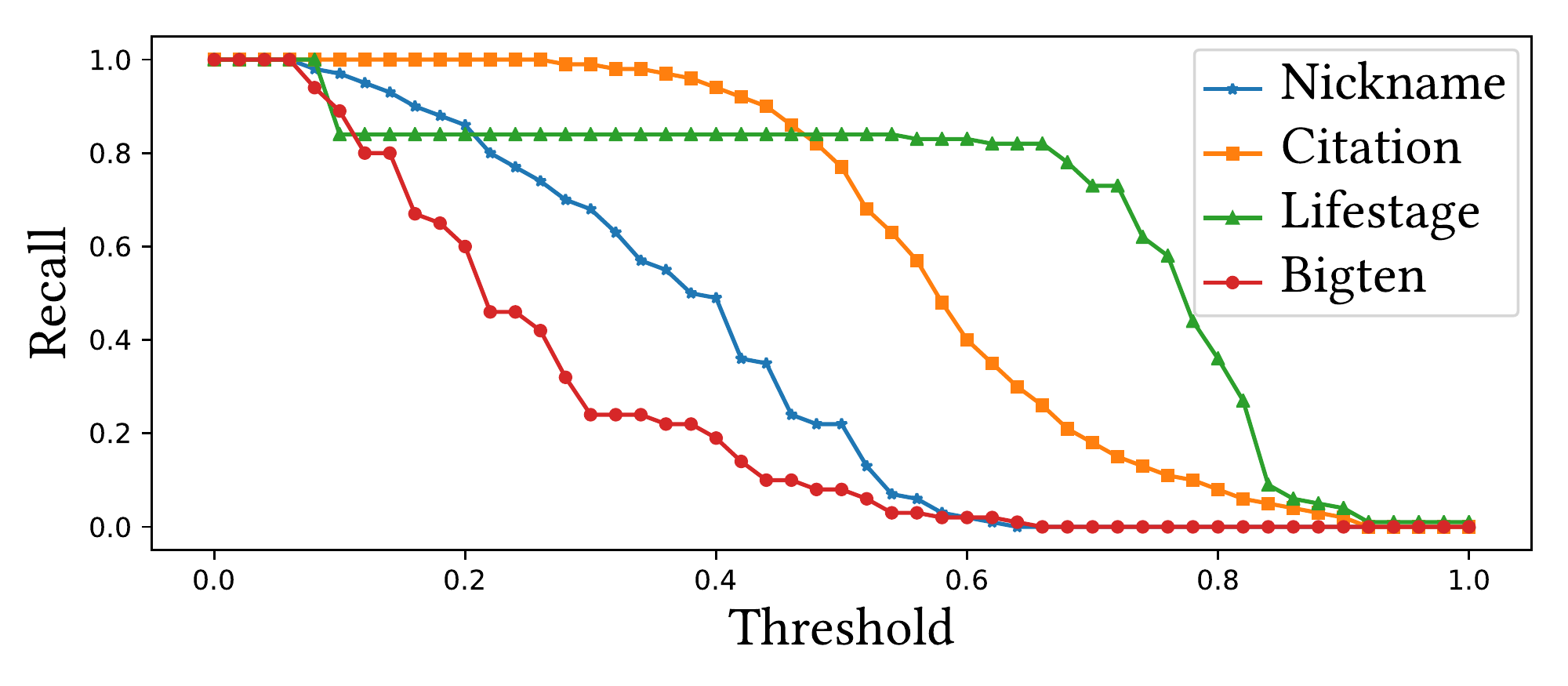} 
	\vspace{-5mm}
	\caption{Recall of TransER for varying threshold $\alpha$.}\label{transerrecall}
	\vspace{-3mm}
\end{figure}

The main problem, as discussed in Section \ref{limsmem}, is that such solutions
cannot guarantee 100\% recall. Consider \transer, which matches
strings using rule $Jaccard(3g(v_i), 3g(v_j)) \geq \alpha$. Assuming a
perfect user $U$ who does not make mistakes when verifying matches,
Figure \ref{transerrecall} shows the recall of \transer\ on our four datasets as we vary
$\alpha$. It shows that to reach 100\% recall, $\alpha$ must be set to
less than 0.08. But that would produce a huge number of matches
(almost the entire Cartesian product), which require a huge amount of
effort from the user to verify. In such cases, it is not difficult to show that
\transer\ would perform worse than \merge.

\paragraph{Comparing with Falcon and Magellan:}
The second way to use current SM/EM solutions is to produce the
matches, then group them into clusters. To examine this approach, we
use \falcon\ \cite{falcon-sigmod17} and \magellan\ \cite{magellan}. A
recent work (name withheld for anonymous reviewing) has adapted
\falcon\ to SM, and shown that it outperforms existing SM
solutions. Thus, {\em \falcon\ can be viewed as a state-of-the-art SM
	solution. \magellan, on the other hand, can be viewed as a
	state-of-the-art EM solution.}  To learn a matcher, both
\falcon\ and \magellan\ require the user to label a set of pairs as
match/non-match. In \magellan\ the user can also debug the matcher to
improve its accuracy.

Once \falcon\ and \magellan\ have produced the matches, we use Markov
clustering in \cite{mcl} to partition the input strings into clusters
that are consistent with the matches. Finally, we ask
one or more users to clean the clusters using \mysplit\ and \merge.

\begin{table}
	\begin{scriptsize}
		\setlength\tabcolsep{2pt}
		\centering
		\begin{tabular}{|l|c|c|c|c|c|c|} \hline
			\begin{tabular}{@{}l@{}}\textbf{Dataset} \end{tabular}& \textbf{1 User} & \textbf{3 Users} & \textbf{5 
				Users} & \textbf{7 Users} & \textbf{9 Users} &\textbf{Labeling} \\ \hline
			Nickname 
			& 1930 (\blue{418})        & 1519 (\blue{629})  & 1210 (\blue{600})  & 958 (\blue{498})   & 807 (\blue{395})   & 14 \\ \hline
			Citation 
			& 1114 (\blue{2})       & 1302 (\blue{874})  & 779 (\blue{501})   & 562 (\blue{350})   & 459 (\blue{282})     & 19 \\ \hline
			Life Stage
			& 22 (\blue{13})          & 21 (\blue{15})       & 21 (\blue{15})       & 22 (\blue{16})       & 22 (\blue{16})
			& 20 \\ \hline
			Big Ten     & 21 (\blue{14})           & 21 (\blue{17})       & 21 (\blue{17})       & 21 (\blue{17})       & 21 (\blue{17})     
			& 20 \\ \hline
		\end{tabular}
	\end{scriptsize}
	\vspace{1mm}\caption{The human times of Falcon.}\label{falconresults}
	\vspace*{-7mm}
\end{table}

\begin{table}
	\begin{scriptsize}
		\setlength\tabcolsep{2pt}
		\centering
		\begin{tabular}{|l|c|c|c|c|c|c|} \hline
			\begin{tabular}{@{}l@{}}\textbf{Dataset} \end{tabular}& \textbf{1 User} & \textbf{3 Users} & \textbf{5 
				Users} & \textbf{7 Users} & \textbf{9 Users} & \begin{tabular}{@{}l@{}}\textbf{Labeling}\\ 
				\textbf{\& Debugging} \end{tabular}\\ \hline
			Nickname 
			& 1482 (\note{-30})      & 1062 (\blue{172})  & 788 (\blue{178})   & 646 (\blue{186})   & 563 (\blue{151})  & 89 \\ \hline
			Citation 
			& 1150 (\blue{38})        & 900 (\blue{472})  & 599 (\blue{321}) & 481 (\blue{269})  & 393 (\blue{216})   & 109 \\ \hline
			Life Stage
			& 85 (\blue{76})           & 85 (\blue{79})      & 85 (\blue{79})        & 85 (\blue{79})       & 85 (\blue{79})
			& 84 \\ \hline
			Big Ten     & 85 (\blue{78})           & 85 (\blue{81})       & 85 (\blue{81})       & 85 (\blue{81})       & 85 (\blue{81})     
			& 84 \\ \hline
		\end{tabular}
	\end{scriptsize}
	\vspace{1mm}\caption{The human times of Magellan.}\label{magellanresults}
	\vspace*{-5mm}
\end{table}

Table \ref{falconresults} shows the human time for \falcon\ on the four datasets. For
example, the first cell ``1930 (418)'' means that for 1 user,
\falcon\ incurs 1930 mins of human time, 418 mins more than
\winston. This time includes the labeling time (14 mins, shown in the
last column). The table shows that \winston\ outperforms \falcon\ in
all cases, reducing human time by 2-874 mins. The larger the dataset,
the more the gain, e.g., more than 14.5 hours on Citation, using 3
users.

Table \ref{magellanresults} shows the human time for \magellan\ on the four datasets.
The meaning of the table cells here are similar to those for \falcon.
The table shows that \winston\ outperforms \magellan\ in all cases,
reducing human time by 38-472 mins, except in the case of 1 user for
Nickname, where it is slower by 30 mins (see the red font).

The above time includes labeling and debugging (the last
column). Interestingly, even if we ignore the labeling and debugging time,
\winston\ {\em still\/} outperforms \magellan\ by a large margin in
all cases requiring 3, 5, 7, and 9 users, for Nickname and
Citation. It is slower only in the case of 1 user, by 119 mins for
Nickname and 71 mins for Citation. Thus, overall \winston\ outperforms
\magellan. In addition, \winston\ is suitable for lay users, whereas
\magellan\ requires the user to have expertise in EM and machine
learning.

A major reason for the worse performance of \falcon\ and \magellan\ is
that they often produce large mixed clusters. For example,
\magellan\ produces clusters of up to 314 strings coming from 137
real-world entities on Nickname, and clusters of up to 98 strings
coming from 62 real-world entities on Citation. Clearly, it is very
time consuming for the user to clean such clusters. In contrast,
\winston\ selects VN plans that produce clusters of only up to 20
strings, which are much easier for the user to understand and clean.
\vspace*{-1mm}\subsection{Additional Experiments}\label{addexpers}

\begin{figure}
	\centering
	\vspace*{-4mm}
	\hspace*{-2mm}
	\includegraphics[height=2.4in]{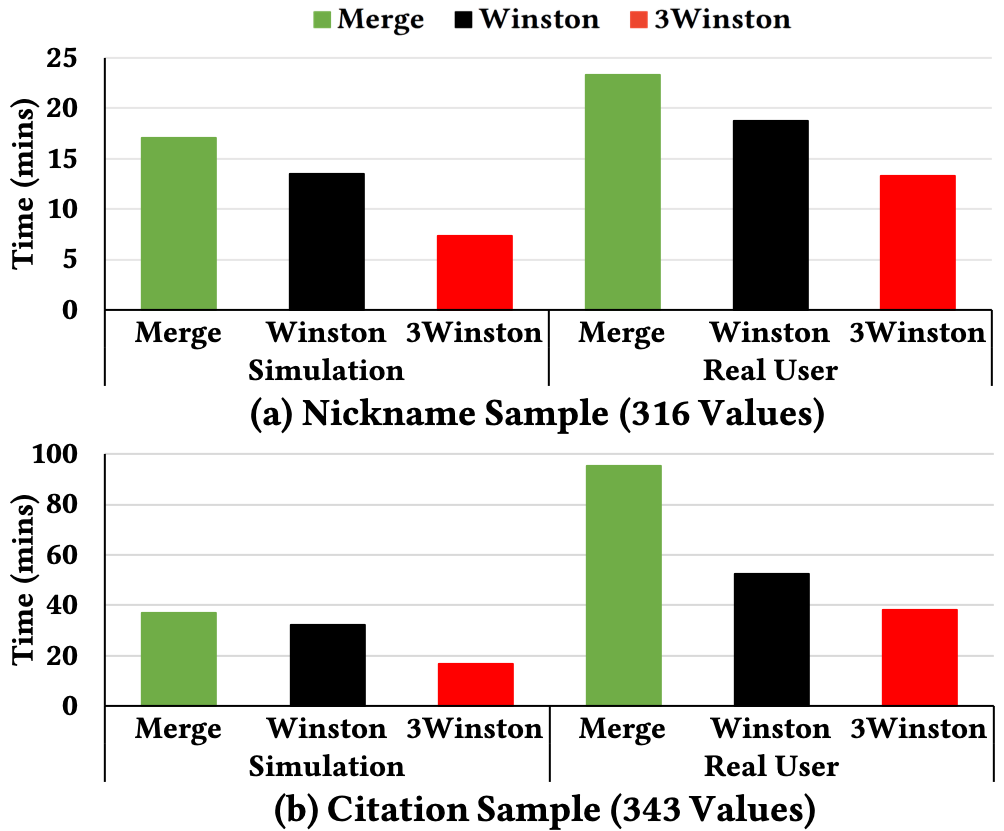} 
	\vspace{-3mm}
	\caption{``Sanity check'' with real users.}\label{sanity}
	\vspace{-1mm}
\end{figure}

\paragraph{``Sanity Check'' with Real Users:} We want to ``sanity check''
our results so far using real users. Extensive checking is very difficult
because it is hard to recruit real users for these time-consuming
experiments.  As a result, we carried out a limited
checking. Specifically, we performed stratified sampling to obtain a
Nickname sample of 316 values and a Citation sample of 343 values. On
each sample we recruited multiple real users and asked them to perform
\merge, \winston\ and {\sf 3Winston} (i.e., \winston\ with 3 users),
taking care to minimize user bias. The right side of Figure
\ref{sanity}.a shows the results for Nickname. For comparison purposes,
the left side of the figure shows the times with synthetic
users. Figure \ref{sanity}.b shows similar results for Citation.

The figures show that ``Simulation'' approximates ``Real User'' quite
well. In both cases, the ordering of the methods is the same. Further,
the results show that \winston\ can do much better than \merge, and
{\sf 3Winston} in turn can do much better than \winston.  While
limited, this result with real users does provide some anecdotal
support for our simulation findings.

\paragraph{Finding Good Plans:} Table \ref{opt} shows that \winston\ finds
good plans. Consider Nickname. Recall that we ran 100
synthetic users for this dataset. For each user $U_i$
\cc\ estimated the costs of all plans then selected plan $p+$, the
one with the least estimated cost. Knowing gold clusters,
however, we can simulate how $U_i$ executes each plan and thus
compute the plan's exact cost.  This allows us to find the rank of
$p+$ on the list of all plans sorted by increasing cost, as
well as the time difference between $p+$ and the best plan. 

The first row of Table \ref{opt} shows this information. Here,
\clang\ considered a space of 100 plans. For all 100 users, it
selected the plan ranked 2nd. The difference between this plan and the
best plan, however, is just 3-4 mins (over 100 users). The
next two cells show the average/min/max times of the best plan, and
the average/min/max difference in percentage. The remaining rows are
similar. Thus, \clang\ did a good job. In many cases,
it selected top-ranked plans, and most importantly, all the selected
plans differ in time from the best plans by only 0-14\% (see the last
column).

\paragraph{Scaling to Large Datasets:} Finally, we examine
how \winston\ scales to large datasets. Table \ref{simdata1} shows the estimated
cleaning time of \merge, \quack, \winston, and {\sf 3Winston},
i.e., \winston\ with 3 users, for synthetic datasets of various
sizes. The table shows that \merge\ is not practical, taking 29 days, 4.4 years,
and 11.5 years for 100K, 500K, and 1M strings, respectively. \quack\ is
better, but still incurs huge times.

\winston, in contrast, can reduce these times drastically, to just 13
days, 9.6 months, and 1.3 years, respectively. As discussed in Section
\ref{intro}, this is because \winston\ provides a better UI, so the
user can do more with less effort. Further, the machine part of
\winston\ outputs clusters that are ``user friendly'', i.e., requiring
little effort for the user to clean.  Finally, \winston\ searches a
large space of plans to find one with minimal estimated human effort.
{\sf 3Winston} does even better, cutting the times to clean 500K and
1M strings to just 2.2 and 3.5 months, respectively. These suggest
that cleaning large datasets with \winston\ indeed can be practical,
especially by dividing the work among multiple users.\vspace*{-1mm}
\begin{table}[t]
	\setlength\tabcolsep{1.5pt}
	\scriptsize
	\centering
	\hspace*{-5mm}
	\begin{tabular}{|l|c|c|c|c|c|} \hline
		\textbf{\begin{tabular}{@{}l@{}} 
				Dataset\end{tabular}} & 
		\textbf{\begin{tabular}{@{}c@{}} Picked Plan \\Rank 
				(Freq)\end{tabular}} & 
		\textbf{\begin{tabular}{@{}c@{}} Size of \\Plan 
				Space\end{tabular}} & 
		\textbf{\begin{tabular}{@{}c@{}} Time Diff\\ to Best 
				Plan\end{tabular}} & 
		\textbf{\begin{tabular}{@{}c@{}} Time 
				of \\Best Plan\end{tabular}} &
		\textbf{\begin{tabular}{@{}c@{}} Diff in \%\end{tabular}} \\ \hline
		\textbf{\begin{tabular}{@{}l@{}} Nickname\end{tabular}} & 
		2 (100) & 100                     & (3,4)                     & 
		\begin{tabular}{@{}c@{}} 2007 (1557,2451)\end{tabular}                    & 
		\begin{tabular}{@{}c@{}} 0.1 (0.1,0.2)\end{tabular}
		\\ \hline
		\textbf{\begin{tabular}{@{}l@{}} Citation\end{tabular}} & 
		1 (100) & 100                     & (0,0)                     & 
		\begin{tabular}{@{}c@{}} 1251 (975,1567)\end{tabular}                    & 
		\begin{tabular}{@{}c@{}} 0 (0,0)\end{tabular}
		\\ \hline
		\begin{tabular}{@{}l@{}} \textbf{Life Stage}\end{tabular} &
		5 (5), 7 (19), 9 (76)                     & 100                     & 
		(0.5,0.7)                     & \begin{tabular}{@{}c@{}} 8.1 
			(7.2,9.2)\end{tabular}                     
		& \begin{tabular}{@{}c@{}} 7 (6,8)\end{tabular}
		\\ \hline
		\textbf{\begin{tabular}{@{}l@{}} Big Ten\end{tabular}} & 
		2 (73)                     & 
		73                     & (0.6,1.2)                     & \begin{tabular}{@{}c@{}} 5.3 
			(5,5.7)\end{tabular}                     
		& \begin{tabular}{@{}c@{}} 14 (9,18)\end{tabular}
		\\ \hline
	\end{tabular}
	\vspace{1mm}
	\caption{The quality of the plans found by Winston.}\label{opt}
	\vspace{-5mm}
\end{table}

\newcommand{\bmerge}{{\sf \textbf{Merge}}}
\newcommand{\bfquack}{{\sf \textbf{Quack}}}

\begin{table}
	\begin{footnotesize}
		\setlength\tabcolsep{2pt}
		\centering
		\begin{tabular}{|c|c|c|c|c|} \hline
			\begin{tabular}{@{}l@{}}\textbf{Dataset} \textbf{Size} \end{tabular}& \bmerge & \bfquack & \bfwinston & 
			\textbf{3\bfwinston} \\ \hline
			10K   & 22           & 22        & 13       & 4       \\ \hline
			100K & 231 (\blue{29d})         & 199      & 107 (\blue{13d})     & 34 (\blue{4.2d})     \\ \hline
			500K &  9415 (\blue{4.4y})     & 3231    & 1688 (\blue{9.6m})   & 387 (\blue{2.2m})   \\ \hline
			1M    &  24449  (\blue{11.5y})  & 19971    & 2710  (\blue{1.3y})   & 618  (\blue{3.5m})   \\ \hline
		\end{tabular}
	\end{footnotesize}
	\vspace{1mm}\caption{Cleaning times vs dataset sizes.}\label{simdata1}
	\vspace*{-4mm}
\end{table}

	\section{Related Work}\label{relatedwork}
\vspace{-1mm}

\paragraph{Data Cleaning:} Data cleaning has received enormous attention
(e.g., \cite{ajax,bigdansing,wisteria,metablocking,das2012automatic,
	distdedup,kolb2011parallel,mozafari2014scaling,aditya2011cidr,marcus2011qurk,
	opt2013park,clamshell,crowddb,wangtransitive2013,activeclean,trifacta,detecterror,
	DBLP:journals/pvldb/DongBHS10,DBLP:journals/debu/FreireCVZ16,DBLP:journals/debu/ArasuCCGKN11,DBLP:journals/debu/ChaudhuriGK06}).
See \cite{dcoverview,qualdc,dcbook,erhardsurvey} for recent tutorials,
surveys, and books. However, as far as we can tell, no published work has
examined the problem of cleaning with 100\% accuracy, as we do for VN
in this paper. {\em Our work here shows that the problem of cleaning to
	reach a desired level of accuracy raises many novel challenges for
	data cleaning.}

\paragraph{Value Normalization:} Much work has addressed VN,
typically under the name ``synonym discovery''. Most solutions use
string/contextual similarities to measure the relatedness of values
\cite{Yates:2009:UMD:1622716.1622724,
	Chakrabarti:2012:FRD:2339530.2339743}, and employ various
techniques, e.g., clustering, regular expressions, learning,
etc. \cite{journals/bmcbi/McCraeC08, Yates:2009:UMD:1622716.1622724}
to match values. However, no work has examined verifying and cleaning
VN results to reach 100\% accuracy, as we do here.

\paragraph{Clustering:} Our work is related to clustering (which we use in
VN). Numerous clustering algorithms exist
\cite{clustersurvey,Xu:2005:SCA:2325810.2327433,
	10.1109/TETC.2014.2330519}, but we are not aware of any work that
has developed a human-driven procedure to clean up clustering output
and tried to minimize the human effort of this procedure. Much work
has also tuned clustering (e.g., \cite{tuningone,tuningtwo}), but for
accuracy. In contrast, our work can be viewed as tuning clustering to
minimize the post-clustering cleaning effort.

\paragraph{String/Entity Matching for the ``Machine'' Part:}
At the core VN is a matching problem, and hence string matching (SM)
and entity matching (EM) solutions can be used in the ``machine''
part.  Numerous such solutions have been developed (e.g., \transer,
\falcon, \magellan, \waldo\ and more \cite{wangtransitive2013,falcon-sigmod17,magellan,waldo}). We
have discussed in Section \ref{limitsol} and experimentally validated in Section \ref{smemsolnlimit}
that these methods do not work well for our context. The main
reason is that they generate large mixed clusters that are very time
consuming for users to clean. {\em This result suggests that when we
	combine a machine part with a human part, it is important to develop
	the machine part such that it generates results that are ``user
	friendly'' for the user in the human part to work with.}

\paragraph{User Interaction Techniques for the ``Human'' Part:}\\
Many recent works on string/entity matching and crowdsourcing solicit
user feedback/action via GUIs to verify and further clean (e.g., {\sf
	CrowdDB}, {\sf CrowdER}, and more
\cite{crowddb,crowder,wangtransitive2013,oeroracle,ercrowderrors,
	waldo}). These works however allow only a limited range of user
actions (e.g., asking users if two tuples match). A recent work,
\waldo\ \cite{waldo}, considers more expressive user actions, such as
showing six values on a single screen and allowing the user to cluster
all six in ``one shot''. The above works differ from \winston\ in two
important ways. First, the range of user actions that they allow is
still quite limited. In contrast, \winston\ considers far more
expressive user actions, such splitting a cluster, merging two
clusters, etc. Second, the above works do not explicitly model the
human effort of the user actions and do not seek to minimize this
total human effort, as \winston\ does. For example, they model the
cost of labeling a value pair or clustering six values to be a fixed
value (e.g., 3 cents paid to a crowd worker), regardless of how much
effort a user puts into doing it. {\em As such, our work can be
	viewed as advancing the recent human-in-the-loop (HILDA) line of
	research, by considering more expressive user actions and studying
	how to optimize their human-effort cost using RDBMS-style
	techniques.}

\paragraph{RDBMS-Style Cleaning Systems:} Many cleaning
works have also adopted an RDBMS-style operator framework, e.g., AJAX
\cite{ajax}, Wisteria \cite{wisteria}, Arnold \cite{arnold}, QuERy
\cite{query}. They however do not consider expressive human
operations, modeling human actions at a coarse level, e.g, labeling a
tuple, converting a dirty tuple into a clean one. In contrast, we
model and estimate the cost of complex human operations, e.g.,
removing a value from a cluster, verifying if a cluster is clean,
etc. Finally, current work typically optimizes for the accuracy and
time of cleaning algorithms (while assuming a ceiling on the human
effort). In contrast, we minimize the human effort, which can be a
major bottleneck in practice.

\paragraph{Interactive Cleaning Systems:} Another
prominent body of work develops interactive cleaning systems (e.g.,
AJAX \cite{ajax}, Potter Wheel \cite{potter}, Wrangler
\cite{Kandel:2011:WIV:1978942.1979444}, Trifacta \cite{trifacta},
ALIAS \cite{aliasfoo}, and \cite{interactivefoo}.  Such systems often
try to maximize cleaning accuracy, or efficiently build data
transformations/cleaning scripts, while minimizing the user effort.
To the best of our knowledge, however, they have not examined the
problem of VN with 100\% accuracy. For example, active learning-based
approaches such as \cite{aliasfoo} do not tell the user what to do (to
reach 100\% accuracy) if after using them the accuracy of the cleaned
dataset is still below 100\%.

\vspace{-1.5mm}\section{Conclusions \& Future Work}\label{conclusions}

We have examined the problem of value normalization with 100\% accuracy. 
We have described \clang, an RDBMS-style solution that
defines human operations, combines them with clustering algorithms to
form hybrid plans, estimates plan costs (in terms of human
verification and cleaning effort), then selects the best plan. 

Overall, our work here shows that it is indeed possible to
apply an RDBMS-style solution approach to the problems of 100\% accurate
cleaning. Going forward, we plan to open source our current
VN solution, explore other clustering algorithms for VN, and explore
applying the solutions here to other cleaning tasks, such as
deduplication, outlier removal, extraction, and data repair.

\bibliographystyle{ACM-Reference-Format}
{
	\scriptsize
	\bibliography{falcon,falcon_new1,relatedwork2,clusteringtuning}


\begin{thebibliography}{49}


\ifx \showCODEN    \undefined \def \showCODEN     #1{\unskip}     \fi
\ifx \showDOI      \undefined \def \showDOI       #1{#1}\fi
\ifx \showISBNx    \undefined \def \showISBNx     #1{\unskip}     \fi
\ifx \showISBNxiii \undefined \def \showISBNxiii  #1{\unskip}     \fi
\ifx \showISSN     \undefined \def \showISSN      #1{\unskip}     \fi
\ifx \showLCCN     \undefined \def \showLCCN      #1{\unskip}     \fi
\ifx \shownote     \undefined \def \shownote      #1{#1}          \fi
\ifx \showarticletitle \undefined \def \showarticletitle #1{#1}   \fi
\ifx \showURL      \undefined \def \showURL       {\relax}        \fi
\providecommand\bibfield[2]{#2}
\providecommand\bibinfo[2]{#2}
\providecommand\natexlab[1]{#1}
\providecommand\showeprint[2][]{arXiv:#2}

\bibitem[\protect\citeauthoryear{??}{ope}{2018}]%
        {openrefine}
 \bibinfo{year}{2018}\natexlab{}.
\newblock \bibinfo{booktitle}{\emph{{OpenRefine} open-source tool.}}
\newblock
\newblock
\shownote{\url{openrefine.org}.}


\bibitem[\protect\citeauthoryear{??}{orc}{2018}]%
        {orclustering}
 \bibinfo{year}{2018}\natexlab{}.
\newblock \bibinfo{booktitle}{\emph{The value normalization capabilities of
  {OpenRefine}.}}
\newblock
\newblock
\shownote{\url{https://github.com/OpenRefine/OpenRefine/wiki/Clustering}.}


\bibitem[\protect\citeauthoryear{Abedjan et~al\mbox{.}}{Abedjan
  et~al\mbox{.}}{2016}]%
        {detecterror}
\bibfield{author}{\bibinfo{person}{Z. Abedjan} {et~al\mbox{.}}}
  \bibinfo{year}{2016}\natexlab{}.
\newblock \showarticletitle{Detecting Data Errors: Where are we and what needs
  to be done?}
\newblock \bibinfo{journal}{\emph{{PVLDB}}} \bibinfo{volume}{9},
  \bibinfo{number}{12} (\bibinfo{year}{2016}), \bibinfo{pages}{993--1004}.
\newblock


\bibitem[\protect\citeauthoryear{Altwaijry et~al\mbox{.}}{Altwaijry
  et~al\mbox{.}}{2015}]%
        {query}
\bibfield{author}{\bibinfo{person}{H. Altwaijry} {et~al\mbox{.}}}
  \bibinfo{year}{2015}\natexlab{}.
\newblock \showarticletitle{QuERy: A Framework for Integrating Entity
  Resolution with Query Processing}.
\newblock \bibinfo{journal}{\emph{PVLDB}} \bibinfo{volume}{9},
  \bibinfo{number}{3} (\bibinfo{year}{2015}), \bibinfo{pages}{120--131}.
\newblock
\showISSN{2150-8097}


\bibitem[\protect\citeauthoryear{Arasu et~al\mbox{.}}{Arasu
  et~al\mbox{.}}{2011}]%
        {DBLP:journals/debu/ArasuCCGKN11}
\bibfield{author}{\bibinfo{person}{A. Arasu} {et~al\mbox{.}}}
  \bibinfo{year}{2011}\natexlab{}.
\newblock \showarticletitle{Towards a Domain Independent Platform for Data
  Cleaning}.
\newblock \bibinfo{journal}{\emph{{IEEE} Data Eng. Bull.}}
  \bibinfo{volume}{34}, \bibinfo{number}{3} (\bibinfo{year}{2011}),
  \bibinfo{pages}{43--50}.
\newblock


\bibitem[\protect\citeauthoryear{Basu et~al\mbox{.}}{Basu
  et~al\mbox{.}}{2002}]%
        {tuningone}
\bibfield{author}{\bibinfo{person}{S. Basu} {et~al\mbox{.}}}
  \bibinfo{year}{2002}\natexlab{}.
\newblock \showarticletitle{Semi-supervised Clustering by Seeding}. In
  \bibinfo{booktitle}{\emph{ICML}}.
\newblock


\bibitem[\protect\citeauthoryear{Bilenko et~al\mbox{.}}{Bilenko
  et~al\mbox{.}}{2004}]%
        {tuningtwo}
\bibfield{author}{\bibinfo{person}{M. Bilenko} {et~al\mbox{.}}}
  \bibinfo{year}{2004}\natexlab{}.
\newblock \showarticletitle{Integrating constraints and metric learning in
  semi-supervised clustering}. In \bibinfo{booktitle}{\emph{ICML}}.
\newblock


\bibitem[\protect\citeauthoryear{Chakrabarti et~al\mbox{.}}{Chakrabarti
  et~al\mbox{.}}{2012}]%
        {Chakrabarti:2012:FRD:2339530.2339743}
\bibfield{author}{\bibinfo{person}{K. Chakrabarti} {et~al\mbox{.}}}
  \bibinfo{year}{2012}\natexlab{}.
\newblock \showarticletitle{A Framework for Robust Discovery of Entity
  Synonyms}. In \bibinfo{booktitle}{\emph{SIGKDD}}.
\newblock


\bibitem[\protect\citeauthoryear{Chaudhuri et~al\mbox{.}}{Chaudhuri
  et~al\mbox{.}}{2006}]%
        {DBLP:journals/debu/ChaudhuriGK06}
\bibfield{author}{\bibinfo{person}{S. Chaudhuri} {et~al\mbox{.}}}
  \bibinfo{year}{2006}\natexlab{}.
\newblock \showarticletitle{Data Debugger: An Operator-Centric Approach for
  Data Quality Solutions}.
\newblock \bibinfo{journal}{\emph{{IEEE} Data Eng. Bull.}}
  \bibinfo{volume}{29}, \bibinfo{number}{2} (\bibinfo{year}{2006}),
  \bibinfo{pages}{60--66}.
\newblock


\bibitem[\protect\citeauthoryear{Chu et~al\mbox{.}}{Chu et~al\mbox{.}}{2016a}]%
        {dcoverview}
\bibfield{author}{\bibinfo{person}{Xu Chu} {et~al\mbox{.}}}
  \bibinfo{year}{2016}\natexlab{a}.
\newblock \showarticletitle{Data Cleaning: Overview and Emerging Challenges}.
  In \bibinfo{booktitle}{\emph{SIGMOD}}.
\newblock


\bibitem[\protect\citeauthoryear{Chu et~al\mbox{.}}{Chu et~al\mbox{.}}{2016b}]%
        {distdedup}
\bibfield{author}{\bibinfo{person}{X. Chu} {et~al\mbox{.}}}
  \bibinfo{year}{2016}\natexlab{b}.
\newblock \showarticletitle{Distributed Data Deduplication}. In
  \bibinfo{booktitle}{\emph{VLDB}}.
\newblock


\bibitem[\protect\citeauthoryear{Chu and Ilyas}{Chu and Ilyas}{2016}]%
        {qualdc}
\bibfield{author}{\bibinfo{person}{X. Chu} {and} \bibinfo{person}{I.~F.
  Ilyas}.} \bibinfo{year}{2016}\natexlab{}.
\newblock \showarticletitle{Qualitative Data Cleaning}.
\newblock \bibinfo{journal}{\emph{{PVLDB}}} \bibinfo{volume}{9},
  \bibinfo{number}{13} (\bibinfo{year}{2016}).
\newblock


\bibitem[\protect\citeauthoryear{Das et~al\mbox{.}}{Das et~al\mbox{.}}{2017}]%
        {falcon-sigmod17}
\bibfield{author}{\bibinfo{person}{S. Das} {et~al\mbox{.}}}
  \bibinfo{year}{2017}\natexlab{}.
\newblock \showarticletitle{Falcon: Scaling Up Hands-Off Crowdsourced Entity
  Matching to Build Cloud Services}. In \bibinfo{booktitle}{\emph{SIGMOD}}.
\newblock


\bibitem[\protect\citeauthoryear{Das~Sarma et~al\mbox{.}}{Das~Sarma
  et~al\mbox{.}}{2012}]%
        {das2012automatic}
\bibfield{author}{\bibinfo{person}{A. Das~Sarma} {et~al\mbox{.}}}
  \bibinfo{year}{2012}\natexlab{}.
\newblock \showarticletitle{An automatic blocking mechanism for large-scale
  de-duplication tasks}. In \bibinfo{booktitle}{\emph{CIKM}}.
\newblock


\bibitem[\protect\citeauthoryear{Dasu and Johnson}{Dasu and Johnson}{2003}]%
        {dcbook}
\bibfield{author}{\bibinfo{person}{T. Dasu} {and} \bibinfo{person}{T.
  Johnson}.} \bibinfo{year}{2003}\natexlab{}.
\newblock \bibinfo{booktitle}{\emph{Exploratory Data Mining and Data
  Cleaning}}.
\newblock \bibinfo{publisher}{John Wiley}.
\newblock
\showISBNx{0-471-26851-8}


\bibitem[\protect\citeauthoryear{Dong et~al\mbox{.}}{Dong
  et~al\mbox{.}}{2010}]%
        {DBLP:journals/pvldb/DongBHS10}
\bibfield{author}{\bibinfo{person}{X. Dong} {et~al\mbox{.}}}
  \bibinfo{year}{2010}\natexlab{}.
\newblock \showarticletitle{Global Detection of Complex Copying Relationships
  Between Sources}.
\newblock \bibinfo{journal}{\emph{{PVLDB}}} \bibinfo{volume}{3},
  \bibinfo{number}{1} (\bibinfo{year}{2010}), \bibinfo{pages}{1358--1369}.
\newblock


\bibitem[\protect\citeauthoryear{Efthymiou et~al\mbox{.}}{Efthymiou
  et~al\mbox{.}}{2015}]%
        {metablocking}
\bibfield{author}{\bibinfo{person}{V. Efthymiou} {et~al\mbox{.}}}
  \bibinfo{year}{2015}\natexlab{}.
\newblock \showarticletitle{Parallel Meta-blocking: Realizing Scalable Entity
  Resolution over Large, Heterogeneous Data}. In \bibinfo{booktitle}{\emph{Big
  Data}}.
\newblock


\bibitem[\protect\citeauthoryear{Fahad et~al\mbox{.}}{Fahad
  et~al\mbox{.}}{2014}]%
        {10.1109/TETC.2014.2330519}
\bibfield{author}{\bibinfo{person}{A. Fahad} {et~al\mbox{.}}}
  \bibinfo{year}{2014}\natexlab{}.
\newblock \showarticletitle{A Survey of Clustering Algorithms for Big Data:
  Taxonomy and Empirical Analysis}.
\newblock \bibinfo{journal}{\emph{IEEE Trans. Emerging Topics in Computing}}
  \bibinfo{volume}{2}, \bibinfo{number}{3} (\bibinfo{year}{2014}),
  \bibinfo{pages}{267--279}.
\newblock
\showISSN{2168-6750}


\bibitem[\protect\citeauthoryear{Firmani et~al\mbox{.}}{Firmani
  et~al\mbox{.}}{2016}]%
        {oeroracle}
\bibfield{author}{\bibinfo{person}{D. Firmani} {et~al\mbox{.}}}
  \bibinfo{year}{2016}\natexlab{}.
\newblock \showarticletitle{Online Entity Resolution Using an Oracle}.
\newblock \bibinfo{journal}{\emph{PVLDB}} \bibinfo{volume}{9},
  \bibinfo{number}{5} (\bibinfo{year}{2016}), \bibinfo{pages}{384--395}.
\newblock
\showISSN{2150-8097}


\bibitem[\protect\citeauthoryear{Franklin et~al\mbox{.}}{Franklin
  et~al\mbox{.}}{2011}]%
        {crowddb}
\bibfield{author}{\bibinfo{person}{M.~J. Franklin} {et~al\mbox{.}}}
  \bibinfo{year}{2011}\natexlab{}.
\newblock \showarticletitle{CrowdDB: answering queries with crowdsourcing}. In
  \bibinfo{booktitle}{\emph{{SIGMOD}}}.
\newblock


\bibitem[\protect\citeauthoryear{Freire et~al\mbox{.}}{Freire
  et~al\mbox{.}}{2016}]%
        {DBLP:journals/debu/FreireCVZ16}
\bibfield{author}{\bibinfo{person}{J. Freire} {et~al\mbox{.}}}
  \bibinfo{year}{2016}\natexlab{}.
\newblock \showarticletitle{Exploring What not to Clean in Urban Data: {A}
  Study Using New York City Taxi Trips}.
\newblock \bibinfo{journal}{\emph{{IEEE} Data Eng. Bull.}}
  \bibinfo{volume}{39}, \bibinfo{number}{2} (\bibinfo{year}{2016}),
  \bibinfo{pages}{63--77}.
\newblock


\bibitem[\protect\citeauthoryear{Galhardas et~al\mbox{.}}{Galhardas
  et~al\mbox{.}}{2001}]%
        {ajax}
\bibfield{author}{\bibinfo{person}{Helena Galhardas} {et~al\mbox{.}}}
  \bibinfo{year}{2001}\natexlab{}.
\newblock \showarticletitle{Declarative Data Cleaning: Language, Model, and
  Algorithms}. In \bibinfo{booktitle}{\emph{VLDB}}.
\newblock


\bibitem[\protect\citeauthoryear{Haas et~al\mbox{.}}{Haas
  et~al\mbox{.}}{2015}]%
        {wisteria}
\bibfield{author}{\bibinfo{person}{D. Haas} {et~al\mbox{.}}}
  \bibinfo{year}{2015}\natexlab{}.
\newblock \showarticletitle{Wisteria: Nurturing Scalable Data Cleaning
  Infrastructure}. In \bibinfo{booktitle}{\emph{VLDB}}.
\newblock


\bibitem[\protect\citeauthoryear{Haas et~al\mbox{.}}{Haas
  et~al\mbox{.}}{2016}]%
        {clamshell}
\bibfield{author}{\bibinfo{person}{D. Haas} {et~al\mbox{.}}}
  \bibinfo{year}{2016}\natexlab{}.
\newblock \showarticletitle{{CLAMShell}: Speeding up Crowds for Low-latency
  Data Labeling}. In \bibinfo{booktitle}{\emph{VLDB}}.
\newblock


\bibitem[\protect\citeauthoryear{He et~al\mbox{.}}{He et~al\mbox{.}}{2016}]%
        {interactivefoo}
\bibfield{author}{\bibinfo{person}{J. He} {et~al\mbox{.}}}
  \bibinfo{year}{2016}\natexlab{}.
\newblock \showarticletitle{Interactive and Deterministic Data Cleaning}. In
  \bibinfo{booktitle}{\emph{{SIGMOD}}}.
\newblock


\bibitem[\protect\citeauthoryear{Heer et~al\mbox{.}}{Heer
  et~al\mbox{.}}{2015}]%
        {trifacta}
\bibfield{author}{\bibinfo{person}{J. Heer} {et~al\mbox{.}}}
  \bibinfo{year}{2015}\natexlab{}.
\newblock \showarticletitle{Predictive Interaction for Data Transformation}. In
  \bibinfo{booktitle}{\emph{CIDR}}.
\newblock


\bibitem[\protect\citeauthoryear{Jain et~al\mbox{.}}{Jain
  et~al\mbox{.}}{1999}]%
        {clustersurvey}
\bibfield{author}{\bibinfo{person}{A.~K. Jain} {et~al\mbox{.}}}
  \bibinfo{year}{1999}\natexlab{}.
\newblock \showarticletitle{Data Clustering: {A} Review}.
\newblock \bibinfo{journal}{\emph{{ACM} Comput. Surv.}} \bibinfo{volume}{31},
  \bibinfo{number}{3} (\bibinfo{year}{1999}), \bibinfo{pages}{264--323}.
\newblock


\bibitem[\protect\citeauthoryear{Jeffery et~al\mbox{.}}{Jeffery
  et~al\mbox{.}}{2013}]%
        {arnold}
\bibfield{author}{\bibinfo{person}{S.~R. Jeffery} {et~al\mbox{.}}}
  \bibinfo{year}{2013}\natexlab{}.
\newblock \showarticletitle{Arnold: Declarative Crowd-Machine Data
  Integration}. In \bibinfo{booktitle}{\emph{{CIDR}}}.
\newblock


\bibitem[\protect\citeauthoryear{Kandel et~al\mbox{.}}{Kandel
  et~al\mbox{.}}{2011}]%
        {Kandel:2011:WIV:1978942.1979444}
\bibfield{author}{\bibinfo{person}{S. Kandel} {et~al\mbox{.}}}
  \bibinfo{year}{2011}\natexlab{}.
\newblock \showarticletitle{Wrangler: Interactive Visual Specification of Data
  Transformation Scripts}. In \bibinfo{booktitle}{\emph{{SIGCHI}}}.
  \bibinfo{pages}{3363--3372}.
\newblock
\showISBNx{978-1-4503-0228-9}


\bibitem[\protect\citeauthoryear{Khayyat et~al\mbox{.}}{Khayyat
  et~al\mbox{.}}{2015}]%
        {bigdansing}
\bibfield{author}{\bibinfo{person}{Z. Khayyat} {et~al\mbox{.}}}
  \bibinfo{year}{2015}\natexlab{}.
\newblock \showarticletitle{{BigDansing}: A System for Big Data Cleansing}. In
  \bibinfo{booktitle}{\emph{SIGMOD}}.
\newblock


\bibitem[\protect\citeauthoryear{Kolb et~al\mbox{.}}{Kolb
  et~al\mbox{.}}{2011}]%
        {kolb2011parallel}
\bibfield{author}{\bibinfo{person}{L. Kolb} {et~al\mbox{.}}}
  \bibinfo{year}{2011}\natexlab{}.
\newblock \showarticletitle{Parallel Sorted Neighborhood Blocking with
  {MapReduce}}. In \bibinfo{booktitle}{\emph{BTW}}.
\newblock


\bibitem[\protect\citeauthoryear{Konda et~al\mbox{.}}{Konda
  et~al\mbox{.}}{2016}]%
        {magellan}
\bibfield{author}{\bibinfo{person}{P. Konda} {et~al\mbox{.}}}
  \bibinfo{year}{2016}\natexlab{}.
\newblock \showarticletitle{Magellan: Toward Building Entity Matching
  Management Systems}.
\newblock \bibinfo{journal}{\emph{{PVLDB}}} \bibinfo{volume}{9},
  \bibinfo{number}{12} (\bibinfo{year}{2016}), \bibinfo{pages}{1197--1208}.
\newblock


\bibitem[\protect\citeauthoryear{Krishnan et~al\mbox{.}}{Krishnan
  et~al\mbox{.}}{2016}]%
        {activeclean}
\bibfield{author}{\bibinfo{person}{S. Krishnan} {et~al\mbox{.}}}
  \bibinfo{year}{2016}\natexlab{}.
\newblock \showarticletitle{ActiveClean: Interactive Data Cleaning For
  Statistical Modeling}.
\newblock \bibinfo{journal}{\emph{{PVLDB}}} \bibinfo{volume}{9},
  \bibinfo{number}{12} (\bibinfo{year}{2016}).
\newblock


\bibitem[\protect\citeauthoryear{Marcus et~al\mbox{.}}{Marcus
  et~al\mbox{.}}{2011}]%
        {marcus2011qurk}
\bibfield{author}{\bibinfo{person}{A. Marcus} {et~al\mbox{.}}}
  \bibinfo{year}{2011}\natexlab{}.
\newblock \showarticletitle{Crowdsourced databases: Query processing with
  people}. In \bibinfo{booktitle}{\emph{CIDR}}.
\newblock


\bibitem[\protect\citeauthoryear{McCrae and Collier}{McCrae and
  Collier}{2008}]%
        {journals/bmcbi/McCraeC08}
\bibfield{author}{\bibinfo{person}{J. McCrae} {and} \bibinfo{person}{N.
  Collier}.} \bibinfo{year}{2008}\natexlab{}.
\newblock \showarticletitle{Synonym set extraction from the biomedical
  literature by lexical pattern discovery.}
\newblock \bibinfo{journal}{\emph{BMC Bioinformatics}}  \bibinfo{volume}{9}
  (\bibinfo{year}{2008}).
\newblock


\bibitem[\protect\citeauthoryear{Miller}{Miller}{1956}]%
        {MillerSTMCapacity}
\bibfield{author}{\bibinfo{person}{G.~A. Miller}.}
  \bibinfo{year}{1956}\natexlab{}.
\newblock \showarticletitle{{The magical number seven plus or minus two: some
  limits on our capacity for processing information.}}
\newblock \bibinfo{journal}{\emph{Psychological Review}} \bibinfo{volume}{63},
  \bibinfo{number}{2} (\bibinfo{year}{1956}).
\newblock
\showISSN{0033-295X}


\bibitem[\protect\citeauthoryear{Mozafari et~al\mbox{.}}{Mozafari
  et~al\mbox{.}}{2014}]%
        {mozafari2014scaling}
\bibfield{author}{\bibinfo{person}{B. Mozafari} {et~al\mbox{.}}}
  \bibinfo{year}{2014}\natexlab{}.
\newblock \showarticletitle{Scaling Up Crowd-Sourcing to Very Large Datasets: A
  Case for Active Learning}. In \bibinfo{booktitle}{\emph{VLDB}}.
\newblock


\bibitem[\protect\citeauthoryear{Parameswaran and Polyzotis}{Parameswaran and
  Polyzotis}{2011}]%
        {aditya2011cidr}
\bibfield{author}{\bibinfo{person}{A.~G. Parameswaran} {and}
  \bibinfo{person}{N. Polyzotis}.} \bibinfo{year}{2011}\natexlab{}.
\newblock \showarticletitle{Answering Queries using Humans, Algorithms and
  Databases}. In \bibinfo{booktitle}{\emph{CIDR}}.
\newblock


\bibitem[\protect\citeauthoryear{Park and Widom}{Park and Widom}{2013}]%
        {opt2013park}
\bibfield{author}{\bibinfo{person}{H. Park} {and} \bibinfo{person}{J. Widom}.}
  \bibinfo{year}{2013}\natexlab{}.
\newblock \showarticletitle{Query Optimization over Crowdsourced Data}. In
  \bibinfo{booktitle}{\emph{VLDB}}.
\newblock


\bibitem[\protect\citeauthoryear{Rahm and Do}{Rahm and Do}{2000}]%
        {erhardsurvey}
\bibfield{author}{\bibinfo{person}{E. Rahm} {and} \bibinfo{person}{H.~H. Do}.}
  \bibinfo{year}{2000}\natexlab{}.
\newblock \showarticletitle{Data Cleaning: Problems and Current Approaches}.
\newblock \bibinfo{journal}{\emph{{IEEE} Data Eng. Bull.}}
  \bibinfo{volume}{23}, \bibinfo{number}{4} (\bibinfo{year}{2000}).
\newblock


\bibitem[\protect\citeauthoryear{Raman and Hellerstein}{Raman and
  Hellerstein}{2001}]%
        {potter}
\bibfield{author}{\bibinfo{person}{V. Raman} {and} \bibinfo{person}{J.~M.
  Hellerstein}.} \bibinfo{year}{2001}\natexlab{}.
\newblock \showarticletitle{Potter's Wheel: An Interactive Data Cleaning
  System}. In \bibinfo{booktitle}{\emph{VLDB}}.
\newblock


\bibitem[\protect\citeauthoryear{Sarawagi et~al\mbox{.}}{Sarawagi
  et~al\mbox{.}}{2002}]%
        {aliasfoo}
\bibfield{author}{\bibinfo{person}{S. Sarawagi} {et~al\mbox{.}}}
  \bibinfo{year}{2002}\natexlab{}.
\newblock \showarticletitle{{ALIAS:} An Active Learning led Interactive
  Deduplication System}. In \bibinfo{booktitle}{\emph{{VLDB}}}.
\newblock


\bibitem[\protect\citeauthoryear{Van~Dongen}{Van~Dongen}{2008}]%
        {mcl}
\bibfield{author}{\bibinfo{person}{S. Van~Dongen}.}
  \bibinfo{year}{2008}\natexlab{}.
\newblock \showarticletitle{Graph Clustering Via a Discrete Uncoupling
  Process}.
\newblock \bibinfo{journal}{\emph{SIAM J. Matrix Anal. Appl.}}
  \bibinfo{volume}{30}, \bibinfo{number}{1} (\bibinfo{year}{2008}).
\newblock


\bibitem[\protect\citeauthoryear{Verroios et~al\mbox{.}}{Verroios
  et~al\mbox{.}}{2017}]%
        {waldo}
\bibfield{author}{\bibinfo{person}{V. Verroios} {et~al\mbox{.}}}
  \bibinfo{year}{2017}\natexlab{}.
\newblock \showarticletitle{Waldo: An Adaptive Human Interface for Crowd Entity
  Resolution}. In \bibinfo{booktitle}{\emph{SIGMOD}}.
\newblock


\bibitem[\protect\citeauthoryear{Verroios and Garcia{-}Molina}{Verroios and
  Garcia{-}Molina}{2015}]%
        {ercrowderrors}
\bibfield{author}{\bibinfo{person}{V. Verroios} {and} \bibinfo{person}{H.
  Garcia{-}Molina}.} \bibinfo{year}{2015}\natexlab{}.
\newblock \showarticletitle{Entity Resolution with crowd errors}. In
  \bibinfo{booktitle}{\emph{{ICDE}}}.
\newblock


\bibitem[\protect\citeauthoryear{Wang et~al\mbox{.}}{Wang
  et~al\mbox{.}}{2012}]%
        {crowder}
\bibfield{author}{\bibinfo{person}{J. Wang} {et~al\mbox{.}}}
  \bibinfo{year}{2012}\natexlab{}.
\newblock \showarticletitle{CrowdER: Crowdsourcing Entity Resolution}.
\newblock \bibinfo{journal}{\emph{PVLDB}} \bibinfo{volume}{5},
  \bibinfo{number}{11} (\bibinfo{year}{2012}), \bibinfo{pages}{1483--1494}.
\newblock
\showISSN{2150-8097}


\bibitem[\protect\citeauthoryear{Wang et~al\mbox{.}}{Wang
  et~al\mbox{.}}{2013}]%
        {wangtransitive2013}
\bibfield{author}{\bibinfo{person}{J. Wang} {et~al\mbox{.}}}
  \bibinfo{year}{2013}\natexlab{}.
\newblock \showarticletitle{Leveraging Transitive Relations for Crowdsourced
  Joins}. In \bibinfo{booktitle}{\emph{SIGMOD}}.
\newblock
\showISBNx{978-1-4503-2037-5}


\bibitem[\protect\citeauthoryear{Xu and Wunsch}{Xu and Wunsch}{2005}]%
        {Xu:2005:SCA:2325810.2327433}
\bibfield{author}{\bibinfo{person}{R. Xu} {and} \bibinfo{person}{D. Wunsch,
  II}.} \bibinfo{year}{2005}\natexlab{}.
\newblock \showarticletitle{Survey of Clustering Algorithms}.
\newblock \bibinfo{journal}{\emph{Trans. Neur. Netw.}} \bibinfo{volume}{16},
  \bibinfo{number}{3} (\bibinfo{year}{2005}), \bibinfo{pages}{645--678}.
\newblock
\showISSN{1045-9227}


\bibitem[\protect\citeauthoryear{Yates and Etzioni}{Yates and Etzioni}{2009}]%
        {Yates:2009:UMD:1622716.1622724}
\bibfield{author}{\bibinfo{person}{A. Yates} {and} \bibinfo{person}{O.
  Etzioni}.} \bibinfo{year}{2009}\natexlab{}.
\newblock \showarticletitle{Unsupervised Methods for Determining Object and
  Relation Synonyms on the Web}.
\newblock \bibinfo{journal}{\emph{J. Artif. Int. Res.}} \bibinfo{volume}{34},
  \bibinfo{number}{1} (\bibinfo{year}{2009}), \bibinfo{pages}{255--296}.
\newblock
\showISSN{1076-9757}


\end{thebibliography}
}

\appendix

\vspace{-1mm}\section{Defining the Human Part}\label{asecplans}
Algorithm \ref{algo:globalmerging} describes the \merge\ and {\sf GlobalMerge} 
procedures ({\sf LocalMerge} has been described in Section \ref{secmerge}). 

\setlength{\algindent}{1em}
\newcommand{\argindent}{4em}
\newcommand{\firstinpindent}{.7em}
\newcommand{\subroutinespace}{3pt}
\newcommand{\hbt}[2]{h$_{#1,\underline{#2}}$}
\begin{algorithm}[t]
	\begin{scriptsize}\flushleft 
		\caption{Merge Phase} 
		\label{algo:globalmerging}
		\textbf{Procedure} Merge($L$)\\
		\textbf{Input:} \hspace*{\firstinpindent}a list of values $L$ representing 
		output clusters of Split phase \\
		\textbf{Output:} a set of clean clusters $C$ of values in $L$
		\begin{algorithmic}[1]
			\State {\underline{LocalMerge}($L$)}
			\State {$L \leftarrow$ consolidated list of values from 
				\underline{LocalMerge} step}
			\State {\textbf{return} GlobalMerge($L$)}
		\end{algorithmic}
		
		\vspace{\subroutinespace}
		
		\textbf{Procedure} GlobalMerge($L$)\\
		\textbf{Input:} \hspace*{\firstinpindent}a list of values $L$ sorted alphabetically\\
		\textbf{Output:} a set of clean clusters $S$ of values in $L$
		\begin{algorithmic}[1]
			\State {\textbf{while} $|L| > 1$ \textbf{do}}
			\Indstate {\textbf{if} $|L| < 3$ \textbf{then} $B \leftarrow [L[1], L[2]]$ 
				\textbf{else} $B \leftarrow [L[1], L[2], L[3]]$}
			\Statex {\-\hspace{1em} // $B$ is the list of values to be displayed on 
				columns}
			\Indstate {\underline{MarkValuesForGlobalMerge}($B, L$)}
			\Statex {\-\hspace{1em} // at the end, user selects ``global merge" 
				button}
			\Indstate {\textbf{for} $i \leftarrow 1, \dots, |B|$ \textbf{do}}
			\Indstate[2] {Merge $B[i]$ and values marked to match it into cluster $s$}
			\Indstate[2] {Remove values in $s$ from $L$, $S \leftarrow S \cup \{s\}$}
			\State {\textbf{return} $S$}
		\end{algorithmic}
		
		\vspace{\subroutinespace}
		
		\textbf{Procedure} \underline{MarkValuesForGlobalMerge}($B, D$)\\
		\textbf{Input:} \hspace*{\firstinpindent}a list $B$ of values on columns, a list $D$ of values on rows\\
		\textbf{Output:} links among values in $B$ and $D$ that match
		\begin{algorithmic}[1]
			\State {\textbf{for each} $i \leftarrow 1, \dots, |B|$ \textbf{do}}
			\Indstate {$b \leftarrow B[i], (\underline{e}, \underline{t}) \leftarrow$ 
				\uamemorize($b$) }
			\Indstate {\textbf{if} $\underline{e}$ is not null \textbf{then} 
				\uafocus(\hbt{b}{t}), \uaselect(\hbt{b}{t})}
			\Statex {\-\hspace{1em} // h$_{x,y}$ is a checkbox to be selected if $x$ 
				and $y$ match}
			\State {\textbf{for each} $j \leftarrow 1, \dots, |D|$ \textbf{do}}
			\Indstate {$d \leftarrow D[j], (\underline{e}, \underline{t}) \leftarrow$ 
				\uarecall($d$) }
			\Indstate {\textbf{if} $\underline{e}$ is not null \textbf{then} 
				\uafocus(\hbt{d}{t}), \uafocus(\hbt{d}{t})}
		\end{algorithmic}
	\end{scriptsize}
\end{algorithm}

\vspace{-1mm}\section{Estimating Plan Costs}\label{aestimatecosts}
In this section we describe the cost estimation formula for various procedures 
used in the human part of value normalization plans and how we have derived 
them.\\

\vspace*{-3mm}\paragraph{SplitCluster Procedure:} To estimate the cost of 
applying \splitcluster \,\,to a cluster $c_i$ during the execution of the plan 
$p_\lambda$ we consider the following three cases:\vspace*{-1mm}\\

\noindent\textit{Case 1} ($\alphalam \geq 0.5$): Recall that when $\alphalam 
\geq 0.5$ we estimate the cost of applying \splitcluster \,\,to $c_i$ as follows:
\begin{equation*}
	\hspace*{-5pt}
	\setlength{\jot}{0.1pt}
	\scriptsize
	\begin{aligned}
		&cost_{SplitCluster}(c_i; \alpha_\lambda \geq 0.5)= 
		\\[-2pt]
		& \sum_{j=1}^{\beta_i} [\underbrace{\rho_p\big( (1 - 
			\alpha_{\lambda})^{j-1} \psi_i 
			,\alpha_{\lambda}\big) + \rho_f + 
			\rho_s}_{q_1}+\underbrace{\rho_d\big( 
			(1 - \alpha_{\lambda})^{j-1} \psi_i\big) + \rho_f + \rho_s}_{q_2} \\[-6pt]
		&+ \underbrace{( (1 - \alpha_{\lambda})^{j-1} \psi_i ) (\rho_f +\rho_m 
			+ (1-\alpha_{\lambda}) \rho_s) }_{q_{3,1}} + \underbrace{\rho_f + 
			\rho_s}_{q_{3,2}} ] \\[-7pt]
	\end{aligned}
\end{equation*}
Also recall that we go through $\beta_i$ iterations of splitting $c_i$ and at 
iteration $j \in \{1, \dots, \beta_i\}$ we split an impure cluster of approximate 
size $(1-\alpha_\lambda)^{j-1} \psi_i$, e.g. at iteration 1 we split the whole 
cluster of size $(1-\alpha_\lambda)^0 \psi_i=\psi_i$. Each iteration 
corresponds 
to a (recursive) call of the \splitcluster. At each execution of \splitcluster, there 
are three lines (numbered 2, 4 and 7 in Algorithm \ref{splitcode}) 
involving user operations and thus only these lines contribute to the cost of the 
procedure.

The cost of line 2 is captured by part $q_1$ of the above formula: it consists of 
the cost of \underline{isPure} (executed on a cluster of size $(1-\alphalam)^{j-1} 
\psi_i$) and then focusing on and selecting ``no" button. Part $q_2$ captures 
the cost of line 4: it consists of the cost of \underline{findDom} and then focusing 
on and selecting ``mark values" button. Finally parts $q_{3,1}$ and $q_{3,2}$ of 
the above formula capture the cost of line 7: $q_{3,1}$ is the cost of 
\underline{{\sf MarkValues}}  and $q_{3,2}$ is the cost of focusing on and 
selecting ``create/clean new cluster" button. Part $q_{3,1}$ in turn consists of 
going through the cluster values (line 3 in \underline{{\sf MarkValues}} pseudo 
code), focusing on each value, matching it with the dominating entity of the 
cluster and selecting the value if they match (i.e. for $1-\alphalam$ fraction of 
the values).\vspace*{-3mm}\\

\noindent\textit{Case 2} ($\alphalam \in [0.1, 0.5)$): We estimate the cost of 
applying \splitcluster \,\,to $c_i$ when $\alphalam \in [0.1, 0.5)$ as follows:
\vspace*{-1mm}
\begin{equation*}
	\scriptsize
	\begin{aligned}
		&cost_{SplitCluster}(c_i; \alphalam \in [0.1, 0.5))= 
		\textstyle\sum_{j=1}^{\beta_i} [\rho_p\big( (1 - \alpha_{\lambda})^{j-1} 
		\psi_i ,\alpha_{\lambda}\big) \\
		&  +\rho_f + \rho_s +\rho_d\big( (1 - \alpha_{\lambda})^{j-1} \psi_i\big) + 
		\rho_f + \rho_s + ( (1 - \alpha_{\lambda})^{j-1} \psi_i )  (\rho_f \\
		& +\rho_m + \alphalam \rho_s)+ \rho_f + \rho_s ]. \\
	\end{aligned}\vspace{-1mm}
\end{equation*}
The derivation is very similar to the previous case. The only 
difference is the fraction of matching values at each execution of \underline{{\sf 
		MarkValues}} which is $\alphalam$ instead of $1-\alphalam$.\vspace*{-3mm}\\

\noindent\textit{Case 3} ($\alphalam < 0.1$): When $\alphalam < 0.1$ we 
estimate the cost of applying \splitcluster \,\,to $c_i$ as follows:
\vspace*{-1mm}
\begin{equation*}
	\hspace*{-5pt}
	\setlength{\jot}{0.01pt}
	\scriptsize
	\begin{aligned}
		&cost_{SplitCluster}(c_i; \alpha_\lambda < 0.1)= 
		\underbrace{\rho_p\big(\psi_i ,\alphalam\big) +\rho_f + \rho_s}_{q'_1} 
		\\[-6pt]
		&+ \underbrace{\rho_d\big(\psi_i\big) + \rho_f + \rho_s}_{q'_2} + 
		\underbrace{\psi_i \rho_z + \psi_i (1 - \tau) (3\rho_f + 2\rho_s) + 		
			\rho_f + \rho_s}_{q'_3} \\[-6pt]
		&+ \textstyle\sum_{j=1}^{\beta_i} 
		[ 3 \rho_z + (3 (1 - \alphalam)^{3(j - 1)} \tau \psi_i - 3) \rho_r \\[5pt]
		&+ (\textstyle\sum_{k=1}^{3} \alphalam (1 - \alphalam)^{3(j - 
			1)+k} \tau \psi_i - 1)(\rho_f + \rho_s) + \rho_f + \rho_s ]\\
	\end{aligned}
\end{equation*}
Here $q'_1$ is the cost of executing \underline{isPure} on $c_i$ and then 
focusing on and selecting ``no" button. $q'_2$ is the cost of 
executing \underline{findDom} and then focusing on and selecting ``clean 
mixed cluster" button. $q'_3$ is the cost of executing \lmerge \-\,\,on $c_i$ 
and the rest of the formula is the cost of executing \gmerge\,\, on the 
results of the previous step. We will describe the costs of \lmerge \,\,and 
\gmerge \,\,in the following sections.\\

\vspace*{-3mm}\paragraph{LocalMerge Procedure:} Recall that for a particular 
plan $p_\lambda$ the split phase result consists of 
approximately $r_\lambda$ pure clusters of input values. Thus the size of the 
input list $L$ to the \lmerge \,\,is $r_\lambda$. User $U$ goes through the 
values in $L$ and for each value, he or she first memorizes it. For $1-\tau$ 
fraction of the values in $L$, $U$ finds a value in his 
or her short-term memory (STM) in which case he or she (1) selects the current 
value, then focuses on and selects the value retrieved from STM and finally 
focuses on and selects the link button. Finally the user focuses on and selects 
``done local merging'' button to proceed to the global merging. Adding up 
these 
costs gives us the cost formula $r_\lambda \rho_z + r_\lambda (1 - \tau) 
(3\rho_f + 2\rho_s) +\rho_f + \rho_s$.\\

\vspace*{-3mm}\paragraph{GlobalMerge Procedure:} \gmerge\ takes as 
input a list 
of values $L$ with approximate size $r'_\lambda$. The \gmerge\ consists of 
possibly several iterations and in each iteration, we 
assume that each of the three values displayed on the columns of the GUI would 
match approximately $\xi r'_\lambda - 1$ values displayed on the rows, 
forming 
clusters of size $\xi r'_\lambda$. Thus the 
number of iterations of \gmerge \,\,would be approximately $g_\lambda = 
\left\lfloor 
r'_\lambda / (3\xi r'_\lambda)\right\rfloor = \left\lfloor 1 / (3\xi) \right\rfloor$.

At iteration $j \in \{1, \dots, g_\lambda\}$ the user sees $r'_\lambda - 3 (j - 1) 
\xi r'_\lambda$ values remained to be matched, three of which are displayed on 
the columns and the rest on the rows of the GUI. The user first memorizes the 
three values on the columns (with total cost of $3 \rho_z$). 
Then for the $r'_\lambda - 3 (j - 1) \xi r'_\lambda - 3$ values on the rows the 
user recalls each value. Lastly for each of the three columns the user focuses on 
and selects checkboxes for $\xi r'_\lambda - 1$ rows. Finally the user focuses on 
and selects ``global merge'' button to finish the current round. Adding up these 
costs would give us the cost formula $\sum_{j=1}^{\left\lfloor 1 / (3\xi) 
	\right\rfloor} [ 3 \rho_z + (r'_\lambda - 3 (j - 1) \xi r'_\lambda - 3)\rho_r + 3 
(\xi r'_\lambda - 1)(\rho_f + \rho_s) + \rho_f + \rho_s ]$\\

\vspace*{-3mm}\paragraph{Estimating the Costs of User Operations:} We now 
describe how 
we estimate the cost $\rho_m$ of the \underline{match} operation, the 
parameters $\gamma$ and $\gamma_0$ of the \underline{isPure} operation 
cost function and the parameters $\eta_1, \eta_2$ and $\eta_3$ of the 
\underline{findDom} operation cost function during the calibration stage.

To estimate $\rho_m$ we first pick three pairs of random values of $V$. We then 
ask the user $U$ to match each pair and, depending on whether they match or 
not, to selects a ``yes'' or ``no'' button. For the $k$th pair we measure the time 
$t_{m,k}$ it takes from when we show the screen containing the pair of values 
and the buttons to $U$ till he or she selects one of the buttons. During this time 
the user matches the values shown on the screen, then focuses on one of the 
buttons and selects it. Hence the time we measure is equal to $\rho_{m, k} + 
\rho_f + \rho_s$ where $\rho_{m,k}$ is our estimated cost of the $k$th 
\underline{match} operation. We then calculate the $\rho_{m,k} = t_{m,k} - 
(\rho_f + \rho_s)$ and estimate $\rho_m$ to be the average of $\rho_{m,k}$s, 
i.e. $\rho_m = \sum_{k=1}^{3} \rho_{m,k}/3$.

For the rest of the parameters above we use the results of HAC(20) we have 
previously run on the input dataset during the calibration phase. Denote the 
results of HAC(20) as $C_{20}$. First we pick three random non-singleton 
clusters $c_1$, $c_2$ and $c_3$ (of different sizes if possible) from $C_{20}$. 
Then we show each $c_k$ and ask $U$ to select a ``yes'' button if $c_k$ is pure 
and a ``no'' button otherwise. We record the time $t_{p,k}$ it takes from when 
we show $c_k$ to $U$ till one of the buttons is selected. We also record which 
button is selected. Using the same timing analysis we described for $\rho_m$, we 
form three equations of the form $\gamma \alpha(c_k) \psi_k + \gamma_0 = 
t_{p,k} - (\rho_f + \rho_s)$ where $k\in \{1,2,3\}$, $\alpha(c_k)$ is the purity of 
$c_k$ and $\psi_k$ is the size of $c_k$. However since we don't know the purity 
of $c_k$s, we use the button $U$ has selected to guess the purity of $c_k$: if $U$ 
has selected the ``yes'' button, we set $\alpha(c_k) = 1$, otherwise we set 
$\alpha(c_k) = a(20)^b$ (we have already estimated $a$ and $b$ during the 
calibration of the purity function). Finally we use ordinary least-squares method 
to solve the system of three equations above to estimate the parameters 
$\gamma$ and $\gamma_0$.

To estimate $\eta_1$ we first pick three clusters $c'_1$, $c'_2$ and $c'_3$ from 
$C_{20}$ such that $\psi'_k = |c'_k| \leq |STM|$. We then show each $c'_k$ to 
$U$ and ask him/her to find the dominating entity of $c'_k$, and then select a 
value in $c'_k$ which refers to dom($c'_k$). For each $c'_k$ we 
measure the time $t_{\eta_1, k}$ it takes from when it is shown to $U$ till 
he/she selects the value referring to dom($c'_k$). Using the same timing analysis 
as above we obtain three equations of the form $t_{\eta_1, k} = \eta_1 \psi'_k + 
\rho_f + \rho_s$. Then we solve each equation for $\eta_1$ and finally average 
the three numbers we get to estimate $\eta_1$ as $\sum_{k=1}^{3} (t_{\eta_1, k} 
- (\rho_f + \rho_s)) / \psi'_k$. 

To estimate $\eta_2$ and $\eta_3$ we follow a similar process: we first pick 
three clusters $c''_1$, $c''_2$ and $c''_3$ from $C_{20}$ such that $\psi''_k = 
|c''_k| > |STM|$. We then show each $c''_k$ to $U$ and ask him/her to find the 
dominating entity of $c''_k$ and then select a value in $c''_k$ which refers to 
dom($c''_k$). For each $c''_k$ we measure the time $t_{\eta_{2,3}, k}$ it takes 
from when it is shown to $U$ till he/she selects the value referring to 
dom($c''_k$). Using the same timing analysis as above we obtain three 
equations 
of the form $t_{\eta_{2,3}, k} = \eta_2 (\psi''_k)^2 + \eta_3 + \rho_f + \rho_s$. 
Finally we use ordinary least-squares method to solve the system of three 
equations above to estimate the parameters $\eta_2$ and $\eta_3$.

\vspace{-3.2mm}\section{Working with Multiple Users}\label{cwinston}
We now describe \cwinston, which extends \winston\ to work with
multiple users.  Assuming $k$ users want to collaborate to normalize a
set $V$ of input values, \cwinston\ goes through four main stages.  In
the first stage, it shows each user a few clusters of values in $V$
and asks them to perform some basic operations on them.
\cwinston\ then uses the results of these operations to tune the
purity function parameters and user operation cost models for each
user (the same way as \winston).  Next, it takes the average of purity
function and cost model parameters to create a single purity function
and a single cost model for each user operation.

In the second stage, \cwinston\ uses the above purity function and
user operation cost models to find the best VN plan. To do so, it uses
the same plan space searching procedure as \winston\ (see Section
\ref{searching}) to find the best plan. 
It then executes the machine part of the best plan to obtain a set 
$\mathcal{C}$ of clusters.

In the third stage, \cwinston\ partitions $\mathcal{C}$ into $k$
subsets of roughly the same number of values.  It then assigns each
subset to one of the users and asks them to clean their respective
subsets of clusters using \mysplit\ and \merge\ algorithms.

In the last stage, \cwinston\ starts by collecting the results of
\mysplit+\merge\ from all the users and for each user, it creates a
list of representative values of the clean clusters he or she has
produced.  It then picks the longest list and divides it into $k$
chunks of roughly the same size.  Next, \cwinston\ asks each user to
merge one of these chunks with the rest of the lists using the
\gmerge\ procedure (see Section \ref{gmerge}).  It then collects the
results from all the users and repeats this stage (i.e., takes
representative values from the merged clusters, divides the largest
list into $k$ chunks, and so on) until all of the lists are
verified/merged.
Algorithm \ref{algorithm:cwinston} shows the pseudocode of \cwinston. 

\paragraph{Estimating the Cost of \bcwinston:} Next, we describe how we estimate the cost 
of \cwinston. To do so, we traverse Algorithm \ref{algorithm:cwinston}, using the same assumption 
as we discussed in Section \ref{estimatecosts}, to arrive at the following cost formula:
\vspace*{-1mm}
\begin{equation*}
	\hspace*{-5pt}
	\setlength{\jot}{0.1pt}
	\scriptsize
	\begin{aligned}
		cost_{\cwinston}{=}&\max_{1\leq i \leq k} \Big(cost_{\mysplit, u_i}(C_i'){+}cost_{\lmerge, u_i}(E_i){+}
		cost_{\gmerge, 
			u_i}(L_i)\Big) \\
		+ &cost_{\text{MultiUserMerge}, U}(S)
	\end{aligned}
	\vspace*{-1mm}
\end{equation*}
where $k$ is the number of users, the $\max$ term finds the longest it takes any of the users $u_i$s 
to perform \mysplit+\merge\ on their respective partitions, and the last term is the cost of multi-user merge 
calculated using the following formula:
\vspace*{-1mm}
\begin{equation*}
	\hspace*{-5pt}
	\setlength{\jot}{0.1pt}
	\scriptsize
	\begin{aligned}
		&cost_{\text{MultiUserMerge}, U}(S)= 
		\sum_{t=1}^{k-1} \Bigg( 
		\max_{u\in U} \Big( 
		\sum_{i=0}^{\frac{|D_t|(1-R_{t-1}\xi)}{3k}} \big(
		3\rho_{z, u} +  \\& \sum_{j=t}^k (3\xi + 1 )(\rho_{f, u} + \rho_{s, u}) + \mu|D_j|(1 - (R_{t-1} + 
		3i)\xi)\rho_{r, u}
		\big)
		\Big)
		\Bigg)
	\end{aligned}
	\vspace*{-1mm}
\end{equation*}
In the above formula, $D_t$ is the largest list of representative values at iteration $t$, 
$R_t$ is the number of entities that are completely merged and removed from the current list $D$ 
after iteration $t$, $\rho_{., u}$s are the cost of human operations for user $u$, $\mu$ is the proportion of the 
rows which, on average, must be examined before all columns are matched, and $\xi$ is described in Section 
\ref{estimatecosts}.

\setlength{\algindent}{1em}
\begin{algorithm}[t]
	\begin{scriptsize}
		\caption{\cwinston}
		\label{algorithm:cwinston}
		
		\begin{flushleft}
			\textbf{Procedure} \cwinston($V,U$) \\
			\textbf{Input: } \hspace*{\firstinpindent}a set $V$ of representative value sets, a set $U = \{u_1, \ldots , 
			u_k\}$ of users\\ 
			\textbf{Output: } a set clean clusters $S$
		\end{flushleft}		
		
		\begin{algorithmic}[1]
			\State {\textbf{for each} $u_i \in U$ \textbf{do}:}
			\Indstate $A_i \gets$ tune purity function and cost model parameters for $u_i$
			
			\State $p_{\lambda^*} \gets$ search plan space to find the best plan using $A_i$s
			\State $C \gets$ run HAC($\lambda^*$) on $V$
			\State Divide $C$ into $C' = \{C_1', C_2', \ldots, C_k'\}$ s.t. $|C_i'| \approx |C| / k$, $S \gets \emptyset$
			\State {\textbf{for each} $u_i \in U$ \textbf{do}: //each user $u_i$ executes \winston\ on $C_i'$}
			\Indstate $D_i \gets$ \mysplit($C_i'$), $E_i \gets$ list of representative values of clusters in $D_i$
			\Indstate \lmerge($E_i$), $L_i \gets$ consolidated list of values from \lmerge\ step
			\Indstate $S \gets S$ $\cup$ \gmerge($L_i$)
			
			\State {\textbf{if} $|U| = 1$ \textbf{then}: \textbf{return} $S$ //effectively, (single-user) \winston}
			\State {\textbf{else}: \textbf{return} MultiUserMerge($S, U$)}
			
		\end{algorithmic} 
		\vspace{\subroutinespace}
		
		\begin{flushleft}
			\textbf{Procedure} MultiUserMerge($S,U$) \\
			\textbf{Input: } \hspace*{\firstinpindent}a set $S$ of clean clusters, a set $U = \{u_1, \ldots , u_k\}$ of 
			users\\ 
			\textbf{Output: } a set clean clusters $S'$ 
		\end{flushleft}
		
		\begin{algorithmic}[1]
			\State $D \gets \emptyset$ //representative values for the sets of clusters in $S$
			\State $S' \gets \emptyset$ //flattened $S$
			\State {\textbf{for each} $S_i \in S$ \textbf{do}:}
			\Indstate {$D_i \gets $ a set of values representing the clusters in $S_i$, $S' \gets S' \cup S_i$}
			
			\State {\textbf{while} each $|D| > 0$ \textbf{do}:}
			\Indstate $D^* \gets \textit{argmax}_{D_j \in D}(|D_j|)$, $D^\dagger \gets D \setminus D^*$, $M \gets 
			\emptyset$ //matches
			\Indstate Divide $D^*$ into $L = \{L_1, L_2, \ldots , L_k\}$ s.t. $|L_i| \approx |D^*| / k$
			
			\Statex \hspace{1.5em}//all users perform merge with their respective column values in parallel
			\Indstate {\textbf{for each} $L_i \in L$ \textbf{do}: $M \gets M$ $\cup$ GroupedMerge($L_i, 
				\textit{copy}(D^\dagger), u_i$)}
			
			\Statex \hspace{1.5em}//resolve matches
			\Indstate {\textbf{for each} $D_j \in D$ \textbf{do}:}
			\Indstate[2] {\textbf{for each} $(v, w) \in M$ \textbf{do}:}
			\Indstate[3] {$D_j \gets D_j \setminus \{v, w\}$}
			\Indstate[3] {Merge the clusters in $S'$ which $v$ and $w$ refer to, and} 
			\Statex {\hspace{5.5em}set $v$ to refer to the new cluster}

			\State \textbf{return } $S'$
			
		\end{algorithmic} 
		\vspace{\subroutinespace}
		
		\begin{flushleft}
			\textbf{Procedure}  GroupedMerge($L, D, u$)\\
			\textbf{Input: } \hspace*{\firstinpindent}a list $L$ of column values, a set $D$ of representative value sets, 
			a user $u$ \\ 
			\textbf{Output: } a set $M$ of matches  
		\end{flushleft}
		
		\begin{algorithmic}[1]
			\State $M \gets \emptyset$
			\State {\textbf{while} each $|L| > 0$ \textbf{do}: //while there are still column values left}
			\Indstate {$r \gets \textit{max}(3, |L|)$, $B \gets r$ values from $L$}
			
			\Indstate {\textbf{for each} $b \in B$ \textbf{do}: \uamemorize$(b)$}
			
			\Indstate {\textbf{for each} $D_j \in D$ \textbf{do}: $M \gets M \cup$ SetMerge($B, D_j, u$)}
			
			\Indstate {$L \gets L \setminus B$}
			
			\State \textbf{return} $M$
			
		\end{algorithmic} 
		\vspace{\subroutinespace}

		\begin{flushleft}
			\textbf{Procedure} SetMerge($B, D_j, u$)\\
			\textbf{Input: } \hspace*{\firstinpindent}a set $B$ of column values, a set $D_j$ of 
			representative values, a user $u$ \\ 
			\textbf{Output: } a set $M'$ of matches 
		\end{flushleft}
		
		\begin{algorithmic}[1]
			\State {$D_j \gets $ sort according to similarity to the values in $B$, $M' \gets \emptyset$}
			
			\State {\textbf{for each} $v \in D_j$ \textbf{do}:}
			\Indstate {\textbf{if} \uarecall$(v)$ \textbf{then}:}
			\Indstate[2] {$M' \gets M' \cup \{(b, v)\}$ s.t. $b \in B \land b $ matches $ v$}
			\Indstate[2] {\textbf{if} $|M| = 3$ \textbf{then}: $\textbf{break}$}
			\Indstate[2] {$D_j \gets D_j \setminus \{v\}$}
			
			\State \textbf{return} $M'$
			
		\end{algorithmic} 
	\end{scriptsize}
\end{algorithm}

Here is how we derive this formula. 
The outer summation corresponds to choosing the longest list of representative values and dividing it among the 
users.
The $\max$ operation chooses the longest it takes any of the users to perform each one of the above iterations, 
i.e., the longest path, which would determine the time it takes the users to collaboratively perform 
MultiUserMerge from start to finish. 

The middle sum corresponds to scanning the remaining lists of representative values each user $u$ has to 
perform at each iteration.
To determine the number of such scans per iteration, we need the number of column values that $u$ has 
to read and memorize for merging. This number is equal to the proportion of the current list $D_t$ still remaining 
to be merged by $u$ which is $\frac{|D_t|(1-R_{t-1}\xi)}{k}$.
Since we show $u$ three column values at a time, we divide the above number by three to arrive at the 
correct number of iterations $\frac{|D_t|(1-R_{t-1}\xi)}{3k}$. 
During each of these scans, $u$ memorizes three column values, hence the $3\rho_{z, u}$ term.

The inner-most sum corresponds to the number of rows examined per each set of three representative values 
during each scan.
At iteration $t$, there are $k - t$ lists of representative values left to appear on the rows. 
Each time $u$ scans one of these lists, he or she matches on average $3\xi$ rows, each of which requires a 
button click.
Additionally, $u$ has to click the merge button, hence the term $(3\xi +1)(\rho_{s, u} + \rho_{f, u})$. 
To account for the number of rows examined in each list before finding the matches, we calculate the number of 
rows left in the list by finding the number of entities removed from the list at the end of this iteration, i.e., 
$|D_j|((R_{t-1} + 3i))\xi$, and then subtracting this value from $|D_j|$. 
We also assume that only a $\mu$ proportion of these rows need to be investigated before finding matches, 
hence the term $\mu|D_j|(1 - (R_{t-1} + 3i)\xi)$. 

\vspace{-1mm}\section{Empirical Evaluation}\label{aeval}
\paragraph{Generating Synthetic Users:} We use a deterministic 
model of a user, i.e. we use constant values for $\rho_f, \rho_s, \rho_m, \rho_r, 
\rho_z, \albr \gamma, \albr \gamma_0, \albr \eta_1, \albr \eta_2, \albr \eta_3$. 
To generate a synthetic user we first assume a constant value for $\rho_f$ and 
$\rho_s$, i.e. $\rho_f = \rho_s = 0.5$. 
We then assume a range of values for each of $\rho_m \in [0.8, 1.2]$, $\rho_r \in [0.3, 0.5]$, 
$\gamma \in [0.1, 0.4]$, $\gamma_0 \in [0.3, 1]$ and $\eta_1 \in [0.2, 0.4]$. 
Next, we generate a random simulated user by uniformly randomly sampling a number 
from each of the above ranges and assigning these values to the corresponding parameters 
of the cost model. 
Finally, we assign the remaining parameters as $\rho_z = 
\rho_r$, $\eta_2 = \eta_1 / (|STM| \times 100)$ and $\eta_3 = 0.99 \eta_1 
|STM|$. 

\end{document}